\documentclass[a4paper,11pt]{article}
\pdfoutput=1 

\usepackage{jheppub} 

\usepackage[T1]{fontenc} 
\usepackage{cancel}

\title{\boldmath Dressing bulk spinor fields in AdS${}_3$}

\author[a]{Gilad Lifschytz}
\author[b]{ and Milan Patra}

\affiliation[a]{
Department of Physics and 
	Haifa Research Center for Theoretical Physics and Astrophysics,\\
	University of Haifa, Haifa 31905, Israel}
	
\affiliation[b]{School of Physical Sciences, National Institute of Science Education and Research\\
An OCC of Homi Bhabha National Institute, Jatani-752050, India}

\emailAdd{giladl@research.haifa.ac.il}
\emailAdd{milan.patra@niser.ac.in}

\abstract{We continue the program of bulk reconstruction for fermionic fields. We reconstruct, from the CFT,  the Dirac fermion field in $AdS_{3}$ coupled to a Chern-Simon gauge field.
We show that the three conditions;  solving the equation of motion, satisfying expected transformation under modular flow and a simple charge distribution at infinity are all compatible and all produce the same bulk operator. We also compute the bulk-boundary tree level three point function from the CFT construction.}

\begin{document} 
\maketitle
\flushbottom

\section{Introduction}
An important question  in  AdS/CFT \cite{Maldacena:1997re} is how to construct the bulk from the boundary CFT.
The program of bulk operator reconstruction is to construct local physics in the bulk from the data in the CFT. Working in the semiclassical limit (large $N$) the question has been answered for free fields \cite{Hamilton:2006az,Kabat:2017mun,Foit:2020} and also for interacting scalar field in $1/N$ approximation \cite{Kabat:2011rz,Kabat:2012av,Kabat:2013wga}, and with other techniques in $AdS_3$ \cite{Anand:2017dav,Chen:2019hdv}.

Recently \cite{Callebaut:2022mqw} a relationship between various ways of constructing the bulk scalar operator from the CFT was explained . In this paper we show that as in the scalar case, these various ways can be implemented for the spinor case and can be used to reconstruct the bulk Dirac fermion coupled to a Chern-Simon gauge field in $AdS_{3}$ to first order in $1/N$. 

The note is organized as follows. In section \ref{background} we will  describe the system we wish to study and the problem  in the CFT 3-point correlator with the zeroth order smearing due to ill-defined $i\epsilon$ prescription. In section \ref{Consistent_conditions} we discuss the consistency of  three possible approaches to reconstruction, namely solving the equation of motion, satisfying expected transformation under modular flow and a simple charge distribution at infinity. As  in the scalar case,  we can reconstruct the bulk operator and cure the  $i\epsilon$ problem order by order in $1/N$ by adding a spectrum of appropriately smeared higher dimension operator. In section \ref{coefficients}, by demanding either of the above  three conditions,  we determine which higher dimension operators one needs as well as determining the coefficients of the higher dimension operators . In section \ref{correlator1} and \ref{correlator2} we explicitly calculate the bulk-boundary 3-point correlator by using the corrected bulk CFT correlator as well as directly, by demanding the cancelation of the  $i\epsilon$ problem, from CFT point of view.

\subsection{Background}\label{background}
We work  in Poincare patch of $\text{AdS}_3$
\begin{equation}
\begin{split}
ds^2=\frac{1}{Z^2}\left(-dt^2+dx^2+dZ^2\right)
\end{split}
\end{equation}
where we have set AdS radius to one. We also define light-front coordinates $x^{\pm}=t\pm x$ and $x_{ij}=x_i-x_j$.\\
If in the CFT we have a primary fermion operator $\psi_{+}$ of dimension $(h,h+\frac{1}{2})$ , then we can build a free bulk Dirac fermion field with mass $m=\Delta-1=2h-\frac{1}{2}$, $\Psi=\begin{pmatrix}\Psi_{-} \\ \Psi_+ \end{pmatrix}$ with \cite{Foit:2020} (For the Euclidean version see \cite{Henningson:1998cd, Mueck:1998iz})

\begin{equation}\label{psi0:branch}
\begin{split}
&\Psi_{+}^{(0)}\left(t, x ,Z\right)=\frac{Z^{\frac{1}{2}}\left(\Delta-\frac{3}{2}\right)}{\pi}\int_{{t^{\prime}}^2+{y^{\prime}}^2<Z^2}dt^{\prime}dy^{\prime}\left(\frac{Z^2-\left(t^{\prime}\right)^2-\left(y^{\prime}\right)^2}{Z}\right)^{\Delta-\frac{5}{2}}\psi_{+}\left(t+t^{\prime},x+i{y^{\prime}}\right)\\
&\Psi_{-}^{(0)}\left(t,x,Z\right)=\frac{Z^{\frac{1}{2}}}{\pi}\int_{{t^{\prime}}^2+{y^{\prime}}^2<Z^2}dt^{\prime}dy^{\prime}\left(\frac{Z^2-\left(t^{\prime}\right)^2-\left(y^{\prime}\right)^2}{Z}\right)^{\Delta-\frac{3}{2}}\partial_{-}\psi_{+}\left(t+t^{\prime},x+i{y^{\prime}}\right)
\end{split}
\end{equation}
 and the near boundary behaviour of the field is
\begin{equation}
\begin{split}
\Psi(t,x,Z)\simeq Z^{\Delta}\begin{pmatrix}0 \\ \psi_+(t,x) \end{pmatrix}+\text{subleading}~~~\text{as $Z \rightarrow 0$}.
\end{split}
\end{equation}

If on the other hand we have a primary fermionic operator $\psi_{-}$ of dimension $(h+\frac{1}{2},h)$, the free bulk field of mass $-m=\Delta-1=2h-\frac{1}{2}$ can be reconstructed as $\Psi=\begin{pmatrix}\Psi_{-} \\ \Psi_+ \end{pmatrix}$, with
\begin{equation}\label{psi0:another_branch}
\begin{split}
\Psi_{+}^{(0)}\left(t,x,Z\right)=&\frac{Z^{\frac{1}{2}}}{\pi}\int_{{\left(t^{\prime}\right)}^2+{\left(y^{\prime}\right)}^2<Z^2}dt^{\prime}dy^{\prime}\left(\frac{Z^2-\left(t^{\prime}\right)^2-\left(y^{\prime}\right)^2}{Z}\right)^{\left(\Delta-\frac{3}{2}\right)} \partial_{+}{\psi_{-}}\left(t+t^{\prime},x+i{y^{\prime}}\right)\\
\Psi_{-}^{(0)}\left(t,x,Z\right)=&\frac{Z^{\frac{1}{2}}\left(\Delta-\frac{3}{2}\right)}{\pi}\int_{{\left(t^{\prime}\right)}^2+{\left(y^{\prime}\right)}^2<Z^2}dt^{\prime}dy^{\prime}\left(\frac{Z^2-\left(t^{\prime}\right)^2-\left(y^{\prime}\right)^2}{Z}\right)^{\left(\Delta-\frac{5}{2}\right)}{\psi_{-}}\left(t+t^{\prime},x+i{y^{\prime}}\right)
\end{split}
\end{equation}
 and the near boundary behaviour of the field is
\begin{equation}
\begin{split}
\Psi(t,x,Z)\simeq Z^{\Delta}\begin{pmatrix}\psi_-(t,x) \\ 0 \end{pmatrix}+\text{subleading}~~~\text{as $Z \rightarrow 0$}.
\end{split}
\end{equation}
In this paper we are interested in a charged Dirac fermion (with mass $m>0$) interacting with a gauge field in $AdS_{3}$. In $AdS_{3}$ the dual of the CFT  conserved current is a $U(1)$ bulk gauge field whose free action is a Chern-Simon term in the bulk \cite{Jensen:2010em}.
The action for interacting Dirac fermion is given by
\begin{equation}
S=\int d^{2+1}x \sqrt{-g} \Bigg[\Big\{\bar{\Psi} (i\Gamma^A e^M_A \mathcal{D}_M -m)\Psi\Big\}-\frac{k}{2}\epsilon^{MLP}A_{M}\partial_{L}A_{P}\Bigg] 
\end{equation}
where, $\bar{\Psi}=\Psi^{\dagger}\Gamma^0$ and the gauge covariant derivative is
\begin{equation}
\begin{split}
\mathcal{D}_M&=\partial_M-\frac{1}{8}\omega_M^{AB}[\Gamma_A,\Gamma_B]+i~\frac{q}{N} A_M.
\end{split}
\end{equation}

Varying the action,  the equation of motion for the fermion is given by
\begin{equation}
\left(i\Gamma^A e^M_A \mathcal{D}_M-m\right)\Psi =0.
\end{equation}
The bulk $2\times2$ Dirac matrices $\Gamma^A$ are given by 
$$\Gamma^{Z}=-i\begin{pmatrix}
  -1 & 0\\\
  0 & 1
\end{pmatrix}~~\text{and}~~\Gamma^a=\begin{pmatrix}
  0 & \sigma^a\\\
  \bar{\sigma}^a & 0
\end{pmatrix}$$
with  $\sigma^a=\left(-i,i\right)$ and $\bar{\sigma}^a=\left(i,i\right)$.\\

In $A_{Z}=0$ gauge and setting $A_{+}=0$ due to the boundary condition for the Chern-Simon term we get the Fermion equation of motion,

\begin{equation}
\begin{split}
&\begin{pmatrix}
  -Z\partial_Z+1-m & 2Z\partial_-\\\\
  -2Z\partial_+ & Z\partial_Z-1-m
\end{pmatrix}\begin{pmatrix}{\Psi_-}\\\\ {\Psi_+}\end{pmatrix}=-\frac{q}{N}Z\begin{pmatrix}2iA_{-}\Psi_{+} \\\\ 0 \end{pmatrix}.
\end{split}
\end{equation}

One can  expand the bulk fields in powers of $\frac{1}{N}$ as
\begin{equation}
\begin{split}
&\Psi=\Psi^{(0)}+\frac{1}{N}\Psi^{(1)}+\cdots\\
&A_M=A^{(0)}_M+\frac{1}{N}A^{(1)}_M+\cdots\\,
\end{split}
\end{equation}
which gives an expansion of  the equation of motion in powers of $\frac{1}{N}$ .
At lowest order the equation of motion for the spinor field, is
\begin{equation}
\begin{split}
\left(i Z\Gamma^a \partial_a+i Z\Gamma^Z \partial_Z-i\Gamma^Z\right)\Psi^{(0)}=m \Psi^{(0)}
\end{split}
\end{equation}
which is already satisfied by \eqref{psi0:branch}.\\
At lowest order the gauge field is given by
\begin{equation}
\begin{split}
A^{(0)}_{+}=0,~~A^{(0)}_{-}=\frac{j_{-}}{k},~~A^{(0)}_{Z}=0.
\end{split}
\end{equation}
where $j_{-}(x^{-})$ is a conserved current in the CFT  with  conformal dimensions $(1,0)$.
Then at first order in $\frac{1}{N}$ we have 

\begin{equation}
\begin{split}
&\begin{pmatrix}
  -Z\partial_Z+1-m & 2Z\partial_-\\\\
  -2Z\partial_+ & Z\partial_Z-1-m
\end{pmatrix}\begin{pmatrix}{\Psi^{(1)}_-}\\\\ {\Psi^{(1)}_+}\end{pmatrix}=-\frac{q Z}{k}\begin{pmatrix}2ij_{-}\left(x^{-}\right)\Psi^{(0)}_{+} \\\\ 0 \end{pmatrix}.
\end{split}
\label{bulkeom}
\end{equation}

\subsubsection{Problem with $\Psi_{\pm}^{(0)}$}\label{ill_defined_corr}

In this paper we choose conventions  such that 
\begin{equation}
<\psi_{+}(x_1) \psi_{+}^{\dagger}(x_{2})>=\frac{1}{(x_{12}^{-})^{2h}(x_{12}^{+})^{2h+1}},
\end{equation}
We also choose the normalization of the 3-point function
\begin{equation}
<\psi_{+}(x_1) \psi_{+}^{\dagger}(x_{2})j_{-}(x_{3})>=\frac{-iq}{2\pi N}\frac{1}{(x_{12}^{-})^{2h-1}(x_{12}^{+})^{2h+1}(x_{13}^{-})(x_{23}^{-})}
\end{equation}
where the order of the operators are with decreasing $i\epsilon$ to the right and $t_{j} \rightarrow t_{j}-i\epsilon_{j}$. With these conventions, one has
\begin{equation}
[j_{-}(x^-), \psi_{+}(y)]=\frac{q}{N}\delta(x^- -y^-)\psi_{+}.
\end{equation}

In appendix  A  we compute the 3-point function (for the case $2h=$ integer)  $\langle\Psi^{(0)}_{\pm}\left(Z, x_1\right)\psi^{\dag}_+\left(x_2\right)j_{-}\left(x_3\right)\rangle$ with the results

\begin{equation}\label{oth:co_p}
\begin{split}
\left<\Psi^{(0)}_+\left(Z, x_1\right)\psi^{\dag}_+\left(x_2\right)j_{-}\left(x_3\right)\right>=\frac{iq}{4h \pi N}\frac{Z^{\frac{1}{2}}\left[(2h-1)\chi_1+\chi_2-2h\chi_1\chi_2\right]}{x_{23}^-x_{12}^+Z^{2h}\left(1-\chi_1\right)^{2h+2}}\left(\frac{\chi_1}{\chi_2}\right)^2\frac{1}{(Y-1)^2}
\end{split}
\end{equation}
where we have introduced the combination
\begin{equation}
  \chi_1=\frac{x_{12}^+x_{12}^-}{Z^2}, \ \  \chi_2=\frac{x_{12}^+x_{13}^-}{Z^2}, \ \ Y=\frac{\chi_2-\chi_1}{\chi_2\left(1-\chi_1\right)}=-\frac{Z^2x_{23}^-}{x_{13}^-\left(x_{12}^+x_{12}^--Z^2\right)}.
\end{equation}

Similarly we have, 

\begin{equation}\label{oth:co_m}
\begin{split}
\left<\Psi^{(0)}_-\left(Z, x_1\right)\psi^{\dag}_+\left(x_2\right)j_{-}\left(x_3\right)\right>=-\frac{iq}{4h \pi N}\frac{Z^{\frac{1}{2}}\left(2h-(2h-1)\chi_2Y\right)}{x_{23}^- Z^{2h+1}(1-\chi_1)^{2h+1}}\left(\frac{\chi_1}{\chi_2}\right)^2\frac{1}{(Y-1)^2}.
\end{split}
\end{equation}

Both expressions suffer from the same problem disccussed in \cite{Kabat:2018pbj, Kabat:2020nvj}. The problem is that there is a pole at $Y=1$, and when $\Psi_{\pm}^{(0)}$ is in the middle of the correlator the $i\epsilon$ prescription inherited from the CFT is ill defined. This  is  becuase the sign of the $i\epsilon$ depends on how exactly we do the crossing of the singularity and on the actual size of $\epsilon_{i}-\epsilon_{j}$ and not just on their sign. This tells us that $\Psi_{\pm}^{(0)}$ is not a good operator in the CFT and must be corrected.

\section{Consistent conditions}\label{Consistent_conditions}

In order to reconstruct the bulk operator one needs some condition. One obvious condition is that the operator will not suffer from the problem described above , rather it will have well defined correlation functions beyond the two point function.  One can try and implement this directly as in \cite{Kabat:2020nvj}, which we do in section 5. Alternatively as explained in \cite{Callebaut:2022mqw}, this is related to having a simple charge distribution condition at infinity written in terms of the bulk fields. In addition a particular charge distribution condition corresponds to a compatible bulk equation of motion and to a compatible transformation under modular flow. 

 In this section we explore three possible conditions: Equation of motion, Charge distribution at infinity, and  transformation under $H_{mod}$ action\footnote{This is a generalization of the approach in \cite{Nakayama:2015mva}, see also \cite{Faulkner:2017vdd}.}. In this section we show that the three conditions we will use are consistent with each other, thus they should reproduce the same bulk operator. In the next section we will reconstruct the bulk operator explicitly and show that indeed the three methods agree.
 
To state the three conditions let us label
\begin{equation}
EOM_{1}=-Z\partial_{Z}\Psi_{-} +(1-m) \Psi_{-} +2Z\partial_{-}\Psi_{+}+\frac{2iq}{Nk}Zj_{-}\Psi_{+}
\end{equation}
\begin{equation}
EOM_{2}=-2Z\partial_{+}\Psi_{-}+Z\partial_{Z}\Psi_{+}-(1+m)\Psi_{+}
\end{equation}
The modular Hamiltonian we will use is the one appropriate to the separation of the CFT into the segment $-R<x<R$ and the rest. This is \cite{Casini:2011kv, deBoer:2016pqk}
\begin{equation}
H_{mod}=\frac{1}{2R}(Q_{0}-R^2P_{0})
\end{equation}
where $Q_{0}$ is the conformal generator of special conformal transformation in the time direction, and $P_{0}$ is the time translation operator.
The three conditions on the bulk fermion operators are then
\begin{enumerate}
\item  Equation of motion
\begin{equation}
EOM_{1}=0, \ \ \ EOM_{2}=0
\end{equation}
\item Charge distribution condition
\begin{equation}
[j_{-}(x^{-}),\Psi_{\pm}(Z,y^{-},y^{+})]=\frac{q}{N}\delta(x^{-}-y^{-})\Psi_{\pm}(Z,y^{-},y^{+})
\label{chdiscon}
\end{equation}
\item Action under $H_{mod}$\footnote{This equation from the bulk point of view is just the bulk Lie derivative, along a Killing vector field, with a compensating gauge transformation restoring $A_{Z}=0$ gauge, reflecting the bulk interpretation \cite{Jafferis:2015del}.}

\begin{equation}
\frac{1}{2\pi i}[H_{mod},\Psi_{\pm}(Z,x,t)]=\xi^{M}\partial_{M} \Psi_{\pm} \pm \frac{x}{2R}\Psi_{\pm} +\frac{Z}{2R}\Psi_{\mp}+\frac{iq}{Nk}\frac{Z^2}{2R}j_{-}\Psi_{\pm}
\label{hmodact}
\end{equation}
where
\begin{equation}
\xi^{\pm}=\frac{1}{2R}(Z^2+(x^{\pm})^{2}-R^2), \ \ \ \xi^{Z}=\frac{t}{R}Z.
\end{equation}

\end{enumerate}

We also have from the CFT\footnote{In our convention $<j_{-}(x^{-})j_{-}(y^{-})>=\frac{k}{2\pi}\frac{1}{(x^{-}-y^{-})^2}$}
\begin{equation}
\frac{1}{2\pi i}[H_{mod},j_{-}]=\frac{1}{2R}(2x^{-}j_{-} +((x^{-})^2-R^2)\partial_{-}j_{-}), \ \ \ [j_{-}(x^{-}),j_{-}(y^{-})]=-ik\partial_{x^{-}}\delta(x^{-}-y^{-}).
\end{equation}

The consistency conditions are then
\begin{equation}
[j_{-}(x^{-}),EOM_{i}(Z,y^{-},y^{+})]=\frac{q}{N}\delta(x^{-}-y^{-})EOM_{i}
\end{equation}
\begin{equation}
\frac{1}{2\pi i}[H_{mod},EOM_1]=(\xi^{M}\partial_{M}-\frac{x}{2R}+\frac{iq}{Nk}\frac{Z^2}{2R}j_{-}) EOM_1 +\frac{Z}{2R} EOM_2 \nonumber
\end{equation}
\begin{equation}
\frac{1}{2\pi i}[H_{mod},EOM_2]=(\xi^{M}\partial_{M}+\frac{x}{2R}+\frac{iq}{Nk}\frac{Z^2}{2R}j_{-}) EOM_2 +\frac{Z}{2R} EOM_1
\end{equation}
and
\begin{equation}
[\Psi_{\pm}, [H_{mod},j_{-}]]+[[H_{mod},\Psi_{\pm}],j_{-}]+[H_{mod},[j_{-}, \Psi_{\pm}]]=0.
\end{equation}

After some algebra one finds that they are obeyed for the above expressions.
Thus all three conditions are consistent and one might expect to find the same bulk operators starting from any one of them, or said differently once one constructs a bulk operator which obeys one of the conditions it will also obey the other two. In the next section we will demonstrate this.

\section{Reconstructing the bulk operator}\label{coefficients}

\subsection{Charge distribution approach}

In this section we solve the charge distribution condition (\ref{chdiscon}) to first order in $\frac{1}{N}$, that is 
\begin{equation}
[j_{-}(x^{-}),(\Psi^{(0)}_{\pm}+\frac{1}{N}\Psi^{(1)}_{\pm})(Z,y^{-},y^{+})]=\frac{q}{N}\delta(x^{-}-y^{-})\Psi^{(0)}_{\pm}(Z,y^{-},y^{+}).
\label{chconfirst}
\end{equation}
This is similar to the approach taken in \cite{Anand:2017dav, Chen:2019hdv}.
Using (\ref{psi0:branch}) and \cite{CarneirodaCunha:2016zmi}
\begin{equation}
\int_{t'^2+y^2\leq Z^2} \left( \frac{Z^2-t'^2-y^2}{Z}\right)^{\Delta-2} {\cal O}(t+t',x+iy)=\pi \Gamma(\Delta-1)\sum_{n=0}^{\infty}\frac{Z^{\Delta+2n}}{\Gamma(n+1)\Gamma(\Delta+n)}(\partial_{-}\partial_{+})^{n} {\cal O}(t,x),
\label{zexpan}
\end{equation}
we have
\begin{equation}
\Psi^{(0)}_{+}=\sum_{n}  C_{r}^{(0,+)} Z^{\Delta+2r} (\partial_{+}\partial_{-})^{r} \psi_{+},
\end{equation}
and
\begin{equation}
\Psi_{-}^{(0)}=\sum_{n}  C_{r}^{(0,-)} Z^{\Delta+2r+1} (\partial_{+}\partial_{-})^{r} \partial_{-}\psi_{+},
\end{equation}
where
\begin{equation}
 C_{r}^{(0,+)}=\frac{\Gamma(\Delta-\frac{1}{2})}{\Gamma(r+1)\Gamma(\Delta+r-\frac{1}{2})}, \ \ \ C_{r}^{(0,-)}=\frac{\Gamma(\Delta-\frac{1}{2})}{\Gamma(r+1)\Gamma(\Delta+r+\frac{1}{2})}.
 \end{equation}
 For the correction term, we write an ansatz
 \begin{equation}
 \Psi_{+}^{(1)}=\sum_{m,n}C_{m n}^{(+)} Z^{2m+2n+\Delta+2}\partial_{-}^{m}j_{-}\partial_{-}^{n}\partial_{+}^{m+n+1}\psi_{+}
 \label{psi1+}
 \end{equation}
  \begin{equation}
 \Psi_{-}^{(1)}=\sum_{m,n}C_{m n}^{(-)} Z^{2m+2n+\Delta+1}\partial_{-}^{m}j_{-}\partial_{-}^{n}\partial_{+}^{m+n}\psi_{+}
 \label{psi1-}
 \end{equation}
 We now insert this into equation (\ref{chconfirst}) (for $\Psi_{+}$)  and compare the coefficients of the power $Z^{2r+\Delta}$ for $r=m+n+1$ . We then get an equation
 \begin{eqnarray}
 & & qC_{r}^{(0,+)}(\partial_{+}\partial_{-})^{r}(\delta(x^{-}-y^{-})\psi_{+})+ik\sum_{m+n=r-1}C_{mn}^{(+)}\partial^{m+1}_{-}\delta(x^{-}-y^{-})\partial_{-}^{n}\partial^{m+n+1}_{+}\psi_{+} \nonumber \\
 & = & qC_{r}^{(0,+)}\delta(x^{-}-y^{-})(\partial_{+}\partial_{-})^{r} \psi_{+}
\end{eqnarray}

This is satisfied if
\begin{equation}
C_{mn}^{(+)}=i\frac{q}{k}C_{m+n+1}^{(0,+)} \frac{\Gamma(m+n+2)}{\Gamma(m+2) \Gamma(n+1)}=i\frac{q}{k}\frac{\Gamma(\Delta-\frac{1}{2})}{\Gamma(m+2)\Gamma(n+1)\Gamma(\Delta+m+n+\frac{1}{2})}.
\label{cnm+}
\end{equation}
One can do a similar thing for $\Psi_{-}$ and we get
\begin{equation}
C_{mn}^{(-)}=i\frac{q}{k}\frac{(m+n+1)\Gamma(\Delta-\frac{1}{2})}{\Gamma(m+2)\Gamma(n+1)\Gamma(\Delta+m+n+\frac{1}{2})}.
\label{cnm-}
\end{equation}

\subsection{$H_{mod}$ action}

The form of the action of the modular Hamiltonian on the bulk fermion operator in equation (\ref{hmodact}), implies that the bulk fermions transform under special conformal transformation on the $x^{-}$ coordinate $L_{1}$ , as if it was a local bulk fermion. The additional piece in (\ref{hmodact}) proportional to $j_{-}$ can only come from the action of $\bar{L}_{1}$ (special conformal transformation on $x^{+}$). What are then the possible building blocks that one can use that obey this condition ?. The CFT operators we have are $j_{-}$ and $\psi_{+}$, from which we can make primary operators $J_{\psi}^{(l)}$ with dimension $(h+l+1,h+\frac{1}{2})$ which are made of  linear combination of operators of the form $\partial^{r}_{-}j_{-}\partial^{l-r}_{-}\psi_{+} $. 
\begin{equation}
J_{\psi}^{(l)}=\sum_{s+r=l} d_{r,s}\partial^{r}_{-}j_{-}\partial^{s}_{-}\psi_{+} 
\end{equation}
where
\begin{equation}
d_{r,s}=\frac{(-1)^{r}}{\Gamma(r+1) \Gamma(s+1)\Gamma (r+2)\Gamma (s+2h)}.
\end{equation}

From this we can make non primary operators of the form $\partial_{+}^{l}J_{\psi}^{(l)}$, whose smearing can contribute to $\Psi^{(1)}_{-}$ and $\partial_{+}^{l+1}J_{\psi}^{(l)}$, whose smearing can contribute to $\Psi^{(1)}_{+}$, as well as $(\partial_{-}\partial_{+})^{m}$ acting on them. We can then ask, which of these operators appropriately smeared will transform under $L_{1}$ as a local bulk fermion.

From equations (\ref{psi0:branch}) and (\ref{psi0:another_branch}) we see that   there are two ways of building a local bulk fermion $\Psi_{\pm}$. If you have a primary operator ${\cal O}_{q,q+\frac{1}{2}}$ of dimension $(q,q+\frac{1}{2})$ and a primary operator ${\cal O}_{q+\frac{1}{2}, q}$ of dimension $(q+\frac{1}{2},q)$ you can build a bulk object that transform under $AdS$ isometries as a bulk fermion  in two ways
\begin{equation}
\Psi_{+} \sim Z^{1/2}\int K_{2q}{\cal O}_{q,q+\frac{1}{2}}\ , \ \ \ \Psi_{+}  \sim Z^{1/2}\int K_{2q+1} \partial_{+} {\cal O}_{q+\frac{1}{2}, q}
\end{equation}
and
\begin{equation}
\Psi_{-} \sim Z^{1/2}\int K_{2q}{\cal O}_{q+\frac{1}{2},q}\ , \ \ \ \Psi_{-}  \sim Z^{1/2}\int K_{2q+1} \partial_{-} {\cal O}_{q,q+\frac{1}{2}}.
\end{equation}
where $K_{\Delta}$ is the scalar smearing function \cite{Hamilton:2006az}, for an operator of dimension $\Delta$. Since the only primary operators available for us  is $J_{\psi}^{(l)}$, we can see that if we want objects that transform under $L_{1}$ as if they are a local bulk fermion\footnote{Since we do not have CFT primary fermionic operators other than $\psi_{+}$, we can not build any bulk object that transform under $AdS$ isometries as a local bulk fermion, other than $\Psi_{\pm}^{(0)}$}  then our options are 
\begin{equation}
\Psi^{(1)}_{+} \sim Z^{1/2}\int K_{2h+2l+2}\partial_{+}^{l+1}J_{\psi}^{(l)},
\end{equation}
and
\begin{equation}
\Psi^{(1)}_{-} \sim Z^{1/2}\int K_{2h+2l+1}\partial_{+}^{l}J_{\psi}^{(l)}, \ \  \Psi^{(1)}_{-} \sim Z^{1/2}\int K_{2h+2l+3}\partial_{-}\partial_{+}^{l+1}J_{\psi}^{(l)}.
\end{equation}
From this we have an ansatz for what the bulk fermion operator can be  built from.

\begin{equation}
\Psi^{(1)}_{+}=\frac{Z^{1/2}}{\pi} \sum_{l=0}^{\infty} a_{l} \int_{t'^2+y^2\leq Z^2} \left( \frac{Z^2-t'^2-y^2}{Z}\right)^{2h+2l}\partial_{+}^{l+1}J_{\psi}^{(l)}(t+t',x+iy)
\label{psi1+n}
\end{equation}
\begin{eqnarray}
\Psi^{(1)}_{-} &=& \frac{Z^{1/2}}{\pi} \sum_{l=0}^{\infty} b_{l} \int_{t'^2+y^2\leq Z^2} \left( \frac{Z^2-t'^2-y^2}{Z}\right)^{2h+2l-1}\partial_{+}^{l}J_{\psi}^{(l)}(t+t',x+iy)\nonumber \\
&+& \frac{Z^{1/2}}{\pi} \sum_{l=0}^{\infty} c_{l} \int_{t'^2+y^2\leq Z^2} \left( \frac{Z^2-t'^2-y^2}{Z}\right)^{2h+2l+1}\partial_{-}\partial_{+}^{l+1}J_{\psi}^{(l)}(t+t',x+iy).\nonumber\\
\label{psi1-n}
\end{eqnarray}

We will now show that one can use the equation describing the transformation of the   bulk fermions  to solve for the bulk operator representing the bulk fermion (i.e find $a_l, b_l, c_l$).
We actually only need one of the equations, which we will take to be the one appropriate  to $\Psi_{-}$ (see (\ref{hmodact}))

\begin{equation}
\frac{1}{2\pi i}[H_{mod},\Psi^{(1)}_{-}(Z,x,t)]=\xi^{M}\partial_{M} \Psi^{(1)}_{-} \pm \frac{x}{2R}\Psi^{(1)}_{-} +\frac{Z}{2R}\Psi^{(1)}_{+}+\frac{iq}{Nk}\frac{Z^2}{2R}j_{-}\Psi^{(0)}_{-}.
\label{hmodact-}
\end{equation}

This transformation can be expanded  in terms of the coefficients $(a_{l},b_{l},c_{l})$ as in equations (\ref{psi1+n}) and (\ref{psi1-n}). In order to use that we need to compute the action of $H_{mod}$ on $\partial_{+}^{l}J_{\psi}^{(l)} $ on $\partial_{+}^{l+1}J_{\psi}^{(l)} $ and on $\partial_{-}\partial_{+}^{l+1}J_{\psi}^{(l)} $.

Given that $J_{\psi}^{(l)}$ is a conformal primary of dimensions $(h+l+1,h+\frac{1}{2})$,  a simple computation gives
\begin{equation}
\frac{1}{2\pi i}[H_{mod}, \partial^{l+1}_{+}J_{\psi}^{(l)}]=P.F_{(h+l+1,h+l+1+\frac{1}{2})} + \frac{1}{2R}(l+1)(2h+l+1)\partial^{l}_{+}J_{\psi}^{(l)}\nonumber
\end{equation}
\begin{equation}
\frac{1}{2\pi i}[H_{mod}, \partial_{-}\partial^{l+1}_{+}J_{\psi}^{(l)}]=\partial_{-}(P.F_{(h+l+1,h+l+1+\frac{1}{2})}) + \frac{1}{2R}(l+1)(2h+l+1)\partial_{-}\partial^{l}_{+}J_{\psi}^{(l)}\nonumber
\end{equation}
\begin{equation}
\frac{1}{2\pi i}[H_{mod}, \partial^{l}_{+}J_{\psi}^{(l)}]=P.F_{(h+l+1,h+l+\frac{1}{2})} + \frac{1}{2R}l(2h+l)\partial^{l-1}_{+}J_{\psi}^{(l)}\nonumber
\end{equation}
where $P.F_{(\tilde{h},\tilde{h}\pm\frac{1}{2})}$ represents the transformation of a primary fermion of dimension $(\tilde{h},\tilde{h}\pm\frac{1}{2})$.

Now the terms that transform like the primary fermion or their $\partial_{-}$ derivative are responsible for the first three terms in equation (\ref{hmodact-}), while the other terms will give the fourth term in equation (\ref{hmodact-}).
The $P.F$ terms in $\Psi_{-}^{(1)}$ (proportional to $b_{l}$, and to $c_{l}$)   contributes to the term  $\frac{Z}{2R}\Psi^{(1)}_{+}$. Note that the term proportional to $b_{l}$ is of the kind (\ref{psi0:another_branch}) while the term proportional to $c_{l}$ is of the kind (\ref{psi0:branch}).  Thus  we get an equation (taking into account the correct normalization)
\begin{equation}
\frac{b_{l}}{2h+2l}+c_{l}(2h+2l+1)=a_{l}.
\label{b+c=a}
\end{equation}

We now turn to the fourth term on the right of equation (\ref{hmodact-}). From the modular Hamiltonian action on $\Psi_{-}^{(1)}$ we get
\begin{eqnarray}
& & \frac{iq}{k}Z^2 j_{-}\Psi_{-}^{(0)} = \frac{Z^{1/2}}{\pi} \sum_{l=0}^{\infty} l(2h+l) b_{l} \int_{t'^2+y^2\leq Z^2} \left( \frac{Z^2-t'^2-y^2}{Z}\right)^{2h+2l-1}\partial_{+}^{l-1}J_{\psi}^{(l)}(t+t',x+iy)\nonumber \\
&+& \frac{Z^{1/2}}{\pi} \sum_{l=0}^{\infty} (l+1)(2h+l+1)c_{l} \int_{t'^2+y^2\leq Z^2} \left( \frac{Z^2-t'^2-y^2}{Z}\right)^{2h+2l+1}\partial_{-}\partial_{+}^{l}J_{\psi}^{(l)}(t+t',x+iy)\nonumber
\label{hmodexpan}
\end{eqnarray}

We can now take this equation and compute on both side its two point function with  $J_{\psi}^{(l)}$ for some specific $l$. Since $<J_{\psi}^{(m)}J_{\psi}^{(l)}>\sim \delta_{ml}$, only one term from each of the sums on the right hand side of equation (\ref{hmodexpan}) will remain. We can then expand in powers of $Z$ using (\ref{zexpan})
to compute the coefficients $b_{l}$ and $c_{l}$.
At the leading order in $Z$ for $l\neq 0$ only the $b_{l}$ term contributes and one gets an equation (after stripping off $\partial_{+}^{l-1}$)
\begin{equation}
\frac{l(2h+l)}{2h+2l} b_{l}<J_{\psi}^{(l)}(x)J_{\psi}^{(l)}(y)>=\frac{iq}{k}\frac{\Gamma(2h)}{\Gamma(l)\Gamma(2h+l)}<j_{-}\partial_{-}^{l}\psi_{+}(x) J_{\psi}^{(l)}(y)>
\end{equation}
One can compute
\begin{equation}
\frac{<j_{-}\partial_{-}^{l}\psi_{+}(x)  J_{\psi}^{(l)}(y)>}{<J_{\psi}^{(l)}(x)J_{\psi}^{(l)}(y)>}=\frac{\Gamma(l+2)\Gamma(l+1)\Gamma(2h+l)\Gamma(2h+l+1)}{\Gamma(2h+2l+1)}
\end{equation}
to get
\begin{equation}
b_{l}=\frac{iq}{k} \frac{\Gamma(2h)\Gamma(l+2)\Gamma(2h+l)}{\Gamma(2h+2l)}
\end{equation}
Since the result makes sense also for $l=0$, we take this to be true for all $l$.

Now we can continue to compute $c_{l}$. Expanding to next order in $Z$ one gets an equation (after stripping off $\partial_{+}^{l}$) 
\begin{equation}
b_{l}\frac{l(2h+l)\Gamma(2h+2l)}{\Gamma(2h+2l+2)}+ c_{l}\frac{(l+1)(2h+l+1)}{2h+2l+2}=\frac{iq}{k} \frac{\Gamma(2h)}{\Gamma(l+1)\Gamma(2h+l+1)}\frac{<j_{-}\partial_{-}^{l+1}\psi_{+}(x) J_{\psi}^{(l)}(y)>}{<J_{\psi}^{(l)}(x)J_{\psi}^{(l)}(y)>}.
\end{equation}
One can compute
\begin{equation}
\frac{<j_{-}\partial_{-}^{l+1}\psi_{+}(x) J_{\psi}^{(l)}(y)>}{<J_{\psi}^{(l)}(x)J_{\psi}^{(l)}(y)>}=\frac{\Gamma(l+1)\Gamma(l+2)\Gamma^{2}(2h+l+1)}{\Gamma(2h+2l+3)}(l(2h+2l+2)+(2h+l+1))
\end{equation}
and from this and the known $b_{l}$ we get
\begin{equation}
c_{l}=\frac{iq}{k} \frac{\Gamma(2h)\Gamma(l+1)\Gamma(2h+l+1)}{\Gamma(2h+2l+2)}.
\end{equation}
 Now we can use equation (\ref{b+c=a}) to get
 \begin{equation}
 a_{l}=\frac{iq}{k}\frac{\Gamma(2h)\Gamma(l+1)\Gamma(2h+l)}{\Gamma(2h+2l+1)}(2h+2l+1).
 \end{equation}
 
We can now compare the expressions (\ref{psi1+n}) and (\ref{psi1-n}) with the coefficients $(a_{l},b_{l},c_{l})$  just computed, to the expressions (\ref{psi1+}) and (\ref{psi1-}), with the coefficients given by (\ref{cnm+}) and (\ref{cnm-}). After some computations one finds that they indeed agree, exhibiting the compatebility of the two conditions.

\subsection{Solving the equations of motion}

Now we will  calculate the correction $\Psi^{(1)}_{\pm}$ by using bulk equation of motion $\Psi^{(1)}_{\pm}$ obeys, 
\begin{equation}
\begin{split}
&\begin{pmatrix}{-Z\partial_Z\Psi^{(1)}_-+\left(1-m\right)\Psi^{(1)}_-+2Z\partial_-\Psi^{(1)}_+}\\\\ {-2Z\partial_+\Psi^{(1)}_-+Z\partial_Z\Psi^{(1)}_+-\left(1+m\right)\Psi^{(1)}_+}\end{pmatrix}=-2\frac{q}{k} Z\begin{pmatrix}ij_{-}\left(x^{-}\right)\Psi^{(0)}_{+} \\\\ 0 \end{pmatrix}
\end{split}
\end{equation}
where the mass of the bulk fermion is given by  $m=2h-\frac{1}{2}$.\\

The correction to $\Psi^{(1)}_-$ can be constructed from the operators $\partial_+^lJ^{(l)}_{\psi}$ and $\left(\partial_+\partial_-\right)\partial_+^lJ^{(l)}_{\psi}$. Then we can write $\Psi^{(1)}_-$ as
\begin{equation}
\begin{split}
\Psi^{(1)}_-(Z,x)=&\frac{Z^{\frac{1}{2}}}{\pi}\sum_{l=0}^{\infty}b_{l}\int_{t^{\prime 2}+y^{\prime 2}\leq Z^2}dt^{\prime}dy^{\prime}\left(\frac{Z^{2}-t^{\prime 2}-y^{\prime 2}}{Z}\right)^{2h+2l-1}\partial_+^lJ^{(l)}_{\psi}\left(t+t^{\prime},x+iy^{\prime}\right)\\
&+\frac{Z^{\frac{1}{2}}}{\pi}\sum_{l=0}^{\infty}c_{l}\int_{t^{\prime 2}+y^{\prime 2}\leq Z^2}dt^{\prime}dy^{\prime}\left(\frac{Z^{2}-t^{\prime 2}-y^{\prime 2}}{Z}\right)^{2h+2l+1}\left(\partial_+\partial_-\right)\partial_+^lJ^{(l)}_{\psi}\left(t+t^{\prime},x+iy^{\prime}\right)\\
=&Z^{2h+\frac{3}{2}}\bigg[\sum_{l=0}^{\infty}b_{l}\Gamma(2h+2l)\sum_{n=0}^{\infty}\frac{Z^{2n+2l}}{\Gamma(n+1)\Gamma(2h+2l+n+1)}\left(\partial_-\partial_+\right)^n\partial^l_+J^{(l)}_{\psi}\\
&+\sum_{l=0}^{\infty}c_{l}\Gamma(2h+2l+2)\sum_{n=0}^{\infty}\frac{Z^{2n+2l+2}}{\Gamma(n+1)\Gamma(2h+2l+n+3)}\left(\partial_-\partial_+\right)^{n+1}\partial^l_+J^{(l)}_{\psi}\bigg]
\end{split}
\end{equation}
where in the last line we have used the following smearing integral formula
\begin{equation}
\begin{split}
\int_{t^{\prime 2}+y^{\prime 2}\leq Z^2}dt^{\prime}dy^{\prime}\left(\frac{Z^{2}-t^{\prime 2}-y^{\prime 2}}{Z}\right)^{\Delta-2}\mathcal{O}(t+t^{\prime},x+iy^{\prime})=\pi~\Gamma(\Delta-1)\sum_{n=0}^{\infty}\frac{Z^{\Delta+2n}}{\Gamma(n+1)\Gamma(\Delta+n)}\left(\partial_-\partial_+\right)^n\mathcal{O}
\end{split}
\end{equation}
To determine $b_{l}$ and $c_{l}$ we will use the bottom part of the equation of motion which reads
\begin{equation}\label{bottom_eom_cl}
\begin{split}
Z\partial_Z\Psi^{(1)}_+-\left(1+m\right)\Psi^{(1)}_+=2Z\partial_+\Psi^{(1)}_-
\end{split}
\end{equation}
The R.H.S of the equation \eqref{bottom_eom_cl} can be computed as
\begin{equation}
\begin{split}
\text{R.H.S}=Z^{\frac{1}{2}}\bigg[&\sum_{l=0}^{\infty}\sum_{n=0}^{\infty}2b_{l}~\Gamma(2h+2l)\frac{Z^{2h+2n+2l+2}}{\Gamma(n+1)\Gamma(2h+2l+n+1)}\left(\partial_-\partial_+\right)^n\partial^{l+1}_+J^{(l)}_{\psi}\\
&+\sum_{l=0}^{\infty}\sum_{n=0}^{\infty}2c_{l}~\Gamma(2h+2l+2)\frac{Z^{2h+2l+2n+4}}{\Gamma(n+1)\Gamma(2h+2l+n+3)}\left(\partial_-\partial_+\right)^n\partial_-\partial^{l+2}_+J^{(l)}_{\psi}\bigg]
\end{split}
\end{equation}
Similarly, the correction $\Psi^{(1)}_+$ can be constructed as
\begin{equation}
\begin{split}
\Psi^{(1)}_+(Z,x)=\frac{Z^{\frac{1}{2}}}{\pi}\sum_{l=0}^{\infty}a_{l}\int_{t^{\prime 2}+y^{\prime 2}\leq Z^2}dt^{\prime}dy^{\prime}\left(\frac{Z^{2}-t^{\prime 2}-y^{\prime 2}}{Z}\right)^{2h+2l}\partial_+^{l+1}J^{(l)}_{\psi}\left(t+t^{\prime},x+iy^{\prime}\right)
\end{split}
\end{equation}
Then L.H.S of the equation \eqref{bottom_eom_cl} can be computed as
\begin{equation}
\begin{split}
\text{L.H.S}=Z^{\frac{1}{2}}\sum_{l=0}^{\infty}\sum_{n=0}^{\infty}\frac{a_{l}~(2l+2n+2)\Gamma(2h+2l+1)}{\Gamma(n+1)\Gamma(2h+2l+n+2)}Z^{2h+2l+2n+2}\left(\partial_-\partial_+\right)^n\partial_+^{l+1}J^{(l)}_{\psi}
\end{split}
\end{equation}

We now take both L.H.S and R.H.S and compute its two point function with $J^{(l)}_{\psi}$ for some specific $l$. Then using the fact that $\langle J^{(l)}_{\psi}~ J^{(n)}_{\psi}\rangle \sim \delta_{ln}$ and then expanding in powers of $Z$ we can compute $b_l$ and $c_l$.\\
At the leading order in powers of $Z$, which is $\mathcal{O}\left(Z^{2h+2l+2}\right)$ we will get the following equation
\begin{equation}
\begin{split}
a_l\frac{(2l+2)\Gamma(2h+2l+1)}{\Gamma(2h+2l+2)}\langle J^{(l)}_{\psi}~\partial_+^{l+1}J^{(l)}_{\psi}\rangle=b_l\frac{2~\Gamma(2h+2l)}{\Gamma(2h+2l+1)}\langle J^{(l)}_{\psi}~\partial_+^{l+1}J^{(l)}_{\psi}\rangle
\end{split}
\end{equation}
which gives,
\begin{equation}\label{b_la_l}
\begin{split}
b_l=a_l\frac{(2h+2l)(l+1)}{(2h+2l+1)}
\end{split}
\end{equation}
At the next order in powers of $Z$, which is $\mathcal{O}\left(Z^{2h+2l+4}\right)$ we will get the following equation
\begin{equation}
\begin{split}
a_l\frac{(2l+4)\Gamma(2h+2l+1)}{\Gamma(2h+2l+3)}\langle J^{(l)}_{\psi}~\partial_-\partial_+^{l+2}J^{(l)}_{\psi}\rangle=&b_l\frac{2~\Gamma(2h+2l)}{\Gamma(2h+2l+2)}\langle J^{(l)}_{\psi}~\partial_-\partial_+^{l+2}J^{(l)}_{\psi}\rangle\\
&+c_l\frac{2~\Gamma(2h+2l+2)}{\Gamma(2h+2l+3)}\langle J^{(l)}_{\psi}~\partial_-\partial_+^{l+2}J^{(l)}_{\psi}\rangle\\
\end{split}
\end{equation}
which gives,
\begin{equation}\label{c_la_l}
\begin{split}
c_l=a_l\frac{(2h+l)}{(2h+2l+1)^2}
\end{split}
\end{equation}
Now we solve for $a_l$. For that we will use the top part of the equation of motion which reads
\begin{equation}\label{eq:top}
\begin{split}
-Z\partial_Z\Psi^{(1)}_-+\left(1-m\right)\Psi^{(1)}_-+2Z\partial_-\Psi^{(1)}_+=-2i\frac{q}{k}~Z~j_{-}\left(x^{-}\right)\Psi^{(0)}_{+} 
\end{split}
\end{equation}
The L.H.S of \eqref{eq:top} can be computed as
\begin{equation}\label{eq:top_lhs}
\begin{split}
\text{L.H.S}=&-Z^{\frac{1}{2}}\sum_{l=0}^{\infty}\sum_{n=0}^{\infty}~b_l~Z^{2h+2l+2n+1}\frac{(4h+2l+2n)\Gamma(2h+2l)}{\Gamma(n+1)\Gamma(2h+2l+n+1)}\left(\partial_-\partial_+\right)^n\partial_+^lJ^{(l)}_{\psi}\\
&-Z^{\frac{1}{2}}\sum_{l=0}^{\infty}\sum_{n=0}^{\infty}~c_l~Z^{2h+2l+2n+3}\frac{(4h+2l+2n+2)\Gamma(2h+2l+2)}{\Gamma(n+1)\Gamma(2h+2l+n+3)}\left(\partial_-\partial_+\right)^n\partial_-\partial_+^{l+1}J^{(l)}_{\psi}\\
&+Z^{\frac{1}{2}}\sum_{l=0}^{\infty}\sum_{n=0}^{\infty}~a_l~Z^{2h+2l+2n+3}\frac{2\Gamma(2h+2l+1)}{\Gamma(n+1)\Gamma(2h+2l+n+2)}\left(\partial_-\partial_+\right)^n\partial_-\partial_+^{l+1}J^{(l)}_{\psi}
\end{split}
\end{equation}
And the R.H.S can be computed as
\begin{equation}\label{eq:top_rhs}
\begin{split}
\text{R.H.S}=&Z^{\frac{1}{2}}\left(-\frac{2iq}{k}\right)\sum_{n=0}^{\infty}\frac{\Gamma(2h)}{\Gamma(n+1)~\Gamma(2h+n)}Z^{2h+2n+1}j_-\left(\partial_-\partial_+\right)^n\psi_+
\end{split}
\end{equation}
Now we again take both L.H.S and R.H.S and compute its two point function with $J^{(l)}_{\psi}$ for some specific $l$. Then using the fact that $\langle J^{(l)}_{\psi}~ J^{(n)}_{\psi}\rangle \sim \delta_{ln}$ and then expanding in powers of $Z$ we can compute $a_l$. At $\mathcal{O}\left(Z^{2h+2l+1}\right)$ we have the following equation
\begin{equation}
\begin{split}
-b_l\frac{(4h+2l)\Gamma(2h+2l)}{\Gamma(2h+2l+1)}\langle J^{(l)}_{\psi}\partial_+^lJ^{(l)}_{\psi}\rangle=-2i\frac{q}{k}\frac{\Gamma(2h)}{\Gamma(l+1)\Gamma(2h+l)}\langle j_- \left(\partial_-\partial_+\right)^l\psi_+J^{(l)}_{\psi}\rangle
\end{split}
\end{equation}
Then after stripping off $\partial_+^l$ and using \eqref{b_la_l} we get
\begin{equation}
\begin{split}
a_l=2i\frac{q}{k}\frac{\Gamma(2h)}{\Gamma(l+1)\Gamma(2h+l)}\frac{\Gamma(2h+2l+2)}{\Gamma(2h+2l+1)(l+1)2(2h+l)}\frac{\langle j_-\partial_-^l\psi_+ J^{(l)}_{\psi}\rangle}{\langle J^{(l)}_{\psi}J^{(l)}_{\psi} \rangle}
\end{split}
\end{equation}
Now one can compute
\begin{equation}
\begin{split}
\frac{\langle j_-\partial_-^l\psi_+ J^{(l)}_{\psi}\rangle}{\langle J^{(l)}_{\psi}J^{(l)}_{\psi} \rangle}=\frac{\Gamma(l+2)\Gamma(l+1)\Gamma(2h+l)\Gamma(2h+l+1)}{\Gamma(2h+2l+1)}
\end{split}
\end{equation}
Using this we have the following expression for $a_l$.
\begin{equation}
\begin{split}
a_l=i\frac{q}{k}\frac{(2h+2l+1)\Gamma(2h)\Gamma(l+1)\Gamma(2h+l)}{\Gamma(2h+2l+1)}
\end{split}
\end{equation}
Using this expression of $a_l$, from \eqref{b_la_l} and \eqref{c_la_l} we have
\begin{equation}
\begin{split}
&b_l=i\frac{q}{k}\frac{\Gamma(2h)\Gamma(l+2)\Gamma(2h+l)}{\Gamma(2h+2l)}\\
&c_l=i\frac{q}{k}\frac{\Gamma(2h)\Gamma(l+1)\Gamma(2h+l+1)}{\Gamma(2h+2l+2)}
\end{split}
\end{equation}
which agrees with the expressions computed from solving the modular flow equations, thus again explicitly showing the compatebility of the conditions described in previous section.
One can also solve the bulk equations of motion directly without using the expected properties under modular Hamiltonian. This is more complicated but can be done, see appendix B  for details.
\section{Computing the corrected 3-point function}\label{correlator1}

In this section we will calculate the bulk 3-point function for the situation where $2h$ is an integer,  using  $\Psi^{(1)}_+$ and $\Psi^{(1)}_-$ in the form of
\begin{equation}
\begin{split}
&\Psi^{(1)}_+=\sum_{M,N=0}^{\infty}C^{(+)}_{MN}Z^{2M+2N+m+3}\partial_-^{M}j_-(x^-)~\partial_-^{N}\partial_+^{M+N+1}\psi_+(x^+,x^-)\\
\end{split}
\end{equation}
with (\ref{cnm+})
\begin{equation}
\begin{split}
C^{(+)}_{MN}=i\frac{q}{k} \frac{(2h-1)!}{\Gamma(2h+M+N+1)(M+1)!(N)!}
\end{split}
\end{equation}
where  $\Delta=2h+\frac{1}{2}$. 
We also have
\begin{equation}
\begin{split}
&\Psi^{(1)}_-=\sum_{M,N=0}^{\infty}C^{(-)}_{MN}Z^{2M+2N+m+2}\partial_-^{M}j_-(x^-)~\partial_-^{N}\partial_+^{M+N}\psi_+(x^+,x^-)\\
\end{split}
\end{equation}
with (\ref{cnm-})
\begin{equation}
\begin{split}
C^{(-)}_{MN}=i\frac{q}{k}\frac{(2h-1)!(M+N+1)}{(2h+M+N)!(M+1)!(N)!}.
\end{split}
\end{equation}
To calculate the bulk correlation function we also note that by large-N factorization we have\footnote{Remember that in our convention $<j_{-}(x^{-})j_{-}(y^{-})>=\frac{k}{2\pi}\frac{1}{(x^{-}-y^{-})^2}$ and $<\psi_{+}(x_1) \psi_{+}^{\dagger}(x_{2})>=\frac{1}{(x_{12}^{-})^{2h}(x_{12}^{+})^{2h+1}}$}
\begin{equation}\label{large_N_factor_charge}
\begin{split}
\left<\bigg(\psi_+\left(x_1\right)j_-\left(x_1\right)\bigg)\psi^{\dag}_+\left(x_2\right)j_-\left(x_3\right)\right>&=\left<\psi_+\left(x_1\right)\psi^{\dag}_+\left(x_2\right)\right>\left<j_-\left(x_1\right)j_-\left(x_3\right)\right>\\
&=\frac{k}{2\pi}\frac{1}{\left(x_{12}^{-}\right)^{2h}\left(x_{12}^{+}\right)^{2h+1}\left(x_{13}^-\right)^2}\\
&=\frac{k}{2\pi}\frac{\partial^{2h-1}_{x_2^-}~\partial^{2h}_{x_2^+}~\partial_{x_3^-}}{\Gamma(2h)\Gamma(2h+1)\Gamma(2)}\frac{1}{x_{12}^{-}x_{12}^{+}x_{13}^{-}}
\end{split}
\end{equation}
Hence, using the expression for $C^{(+)}_{MN}$ and \eqref{large_N_factor_charge} the correlator of $\Psi^{(1)}_+$ with $\psi^{\dag}_{+}$ and $j_{-}$ can be calculated as
\begin{equation}
\begin{split}
\left<\Psi^{(1)}_{+}\left(x_1,Z\right)\psi^{\dag}_{+}\left(x_2\right)j_{-}\left(x_3\right)\right>&=\frac{k}{2\pi}\sum_{M,N=0}^{\infty}C^{(+)}_{MN}Z^{2M+2N+m+3}\frac{\partial_{x_1^+}\partial^{2h+N-1}_{x_2^-}~\partial^{2h+M+N}_{x_2^+}~\partial_{x_3^-}^{M+1}}{\Gamma(2h)\Gamma(2h+1)\Gamma(2)}\frac{1}{x_{12}^{-}x_{12}^{+}x_{13}^{-}}\\
&=-i\frac{q}{2 \pi}Z^{\frac{1}{2}}\sum_{M,N=0}^{\infty}\frac{(2h+M+N+1)(2h+N-1)!}{N!(2h)!Z^{2h+2}}\frac{1}{(\chi_1)^{2h+N}(\chi_2)^{M+2}}\\
&=-i\frac{q}{2 \pi}\frac{Z^{\frac{1}{2}}}{Z^{2h+2}}\frac{\left[\chi_2(\chi_1-1)+2h\chi_1(\chi_2-1)\right]}{(2h)\chi_2\left(\chi_1-1\right)^{2h+1}\left(\chi_2-1\right)^2}\\
&=i\frac{q}{2 \pi}\frac{Z^{\frac{1}{2}}}{(2h)Z^{2h+2}}\frac{\left[2h(Y-1)-1\right]}{(\chi_1-1)^{2h}(\chi_2-1)^2}\\
&=i\frac{q}{2 \pi}\frac{Z^{\frac{1}{2}}}{(2h)Z^{2h+2}}\frac{\left[2h(Y-1)-1\right]}{(\chi_1-1)^{2h+2}}\left(\frac{\chi_1}{\chi_2}\right)^2\frac{1}{(Y-1)^2}\\
\end{split}
\end{equation}
which also has  the problematic pole  at $Y=1$. 
Let's now combine this first order correlator with the lowest order correlator, which has the following form
\begin{equation}
\begin{split}
\langle\Psi^{(0)}_+\left(Z, x_1\right)\psi^{\dag}_+\left(x_2\right)j_{-}\left(x_3\right)\rangle&=\frac{iq}{2\pi N}\frac{Z^{\frac{1}{2}}\left[(2h-1)\chi_1+\chi_2-2h\chi_1\chi_2\right]}{2h(x_{23}^-)(x_{12}^+)(Z)^{2h}\left(\chi_1 -1\right)^{2h+2}}\left(\frac{\chi_1}{\chi_2}\right)^2\frac{1}{(Y-1)^2}\\
&=-\frac{iq}{2\pi N}\frac{Z^{\frac{1}{2}}}{2 h Z^{2h+2}}\frac{\left[2h(Y-1)-Y\right]}{(\chi_1-1)^{2h+2}Y}\left(\frac{\chi_1}{\chi_2}\right)^2\frac{1}{(Y-1)^2}
\end{split}
\end{equation}
and then we have
\begin{equation}\label{corrected_correlation_psipn}
\begin{split}
&\langle\left(\Psi^{(0)}_+\left(Z, x_1\right)+\frac{1}{N}\Psi^{(1)}_{+}\left(Z,x_1\right)\right)\psi^{\dag}_+\left(x_2\right)j_{-}\left(x_3\right)\rangle=i\frac{q}{2 \pi N}\frac{1}{Z^{2h+2}}\frac{Z^{\frac{1}{2}}}{Y(\chi_1-1)^{2h+2}}\left(\frac{\chi_1}{\chi_2}\right)^2
\end{split}
\end{equation}
Now as expected the corrected correlation function has no pole at $Y=1$, and so no $i\epsilon$ problem.

Similarly, using the expression for $C^{(-)}_{MN}$ and \eqref{large_N_factor_charge} the correlator of $\Psi^{(1)}_-$ with $\psi^{\dag}_{+}$ and $j_{-}$ can be calculated as
\begin{equation}
\begin{split}
\langle\Psi^{(1)}_{-}\left(x_1,Z\right)\psi^{\dag}_{+}\left(x_2\right)j_{-}\left(x_3\right)\rangle&=\frac{k}{2\pi} Z^{\frac{1}{2}}\sum_{M,N=0}^{\infty}C^{(-)}_{MN}Z^{2h+2M+2N+1}\frac{\partial^{2h+N-1}_{x_2^-}~\partial^{2h+M+N}_{x_2^+}~\partial_{x_3^-}^{M+1}}{(2h)!(2h-1)!}\frac{1}{x_{12}^{-}x_{12}^{+}x_{13}^{-}}\\
&=\ i\frac{q}{2\pi}\frac{Z^{\frac{1}{2}}}{(Z)^{2h+1}}\sum_{M,N=0}^{\infty}\frac{(M+N+1)(2h+N-1)!}{N!(2h)!(x_{13}^-)}\frac{1}{\left(\chi_1\right)^{2h+N}\left(\chi_2\right)^{M+1}}\\
&=i\frac{q}{2\pi}\frac{Z^{\frac{1}{2}}}{(Z)^{2h+1}(x_{13}^-)}\frac{\left[2h(\chi_2-1)+\chi_2(\chi_1-1)\right]}{(2h)(\chi_1-1)^{2h+1}(\chi_2-1)^2}\\
&=-i\frac{q}{2\pi}\frac{Z^{\frac{1}{2}}\left[-2hY+2h+\chi_1\right]}{(2h)(Z)^{2h+1}(x_{23}^-)(\chi_1-1)^{2h+1}}\left\{\frac{Y\chi_2}{\chi_1}\right\}\left(\frac{\chi_1}{\chi_2}\right)^2\frac{1}{(Y-1)^2}
\end{split}
\end{equation}

Again, let's now combine this first order correlator with the lowest order correlator, which has the following form
\begin{equation}
\begin{split}
\langle\Psi^{(0)}_-\left(Z, x_1\right)\psi^{\dag}_+\left(x_2\right)j_{-}\left(x_3\right)\rangle=\frac{iq}{4 h \pi N}\frac{Z^{\frac{1}{2}}\left(2h-(2h-1)\chi_2Y\right)}{x_{23}^- Z^{2h+1}(\chi_1 -1)^{2h+1}}\left(\frac{\chi_1}{\chi_2}\right)^2\frac{1}{(Y-1)^2}
\end{split}
\end{equation}
and then using
\begin{equation}
\begin{split}
\chi_2Y=\frac{\chi_1Y}{1+\chi_1Y-Y}
\end{split}
\end{equation}
we have
\begin{equation}\label{corrected_correlation_psimn}
\begin{split}
\langle\left(\Psi^{(0)}_-\left(Z, x_1\right)+\frac{1}{N}\Psi^{(1)}_{-}\left(Z,x_1\right)\right)\psi^{\dag}_+\left(x_2\right)j_{-}\left(x_3\right)\rangle =i\frac{q}{2\pi N}\frac{Z^{\frac{1}{2}}}{Z^{2h+1}x_{23}^-(\chi_1-1)^{2h+1}}\left(\frac{\chi_1}{\chi_2}\right)
\end{split}
\end{equation}
Now as expected the corrected correlation function has no pole at $Y=1$. One can check that these expressions obey the bulk equation of motion (\ref{bulkeom}).
\section{Computing by canceling the pole}\label{correlator2}
Here we follow a similar line of thought as in \cite{Kabat:2020nvj}.
The zeroth order correlator for $\Psi^{(0)}_+$ is given by
\begin{equation}\label{zerothpsi+}
\begin{split}
\left<\Psi^{(0)}_+\left(Z, x_1\right)\psi^{\dag}_+\left(x_2\right)j_{-}\left(x_3\right)\right>=\frac{\hat{\gamma}Z^{\frac{1}{2}}\left[(2h-1)\chi_1+\chi_2-2h\chi_1\chi_2\right]}{(x_{23}^-)(x_{12}^+)(Z)^{2h}\left(\chi_1 -1\right)^{2h+2}}\left(\frac{\chi_1}{\chi_2}\right)^2\frac{1}{(Y-1)^2}
\end{split}
\end{equation}
with $\hat{\gamma}=\frac{iq}{4h\pi N}$.

For our purposes it is more convinient to note that
\begin{equation}
\begin{split}
(2h-1)\chi_1+\chi_2-2h\chi_1\chi_2=\chi_2\left(\chi_1-1\right)\left[2h(Y-1)-Y\right]
\end{split}
\end{equation}
so
\begin{equation}
\begin{split}
\left<\Psi^{(0)}_+\left(Z, x_1\right)\psi^{\dag}_+\left(x_2\right)j_{-}\left(x_3\right)\right>=\frac{\hat{\gamma}Z^{\frac{1}{2}}{\chi_2\left(\chi_1-1\right)}}{(x_{23}^-)(x_{12}^+)(Z)^{2h}\left(\chi_1 -1\right)^{2h+2}}\left(\frac{\chi_1}{\chi_2}\right)^2\frac{1}{(Y-1)^2}\left[2h(Y-1)-Y\right]
\end{split}
\label{zerothpsi+bet}
\end{equation}
From (\ref{psil+}), the correlator for the higher dimensional operators is
\begin{equation}\label{hpsi+}
\begin{split}
\left<\Psi^{(l)}_{+}\psi^{\dag}_{+}j_{-}\right>=\frac{\hat{\alpha}^{(l)}Z^{\frac{1}{2}}{\chi_2\left(\chi_1-1\right)}}{\left(x_{23}^-\right)\left(x_{12}^+\right)\left(Z\right)^{2h}\left(\chi_1 -1\right)^{2h+2}}\left(\frac{\chi_1}{\chi_2}\right)^2\frac{1}{(Y-1)^2}(-Y)^{l+1}F\left(2h+l,l,2h+2l+2,Y\right)
\end{split}
\end{equation}
So these are our building blocks for cancelling the pole.  The most general expression we can have is 
\begin{equation}\label{perturbativeeq}
\begin{split}
\Psi^{(1)}_{+}=\sum_{l=0}^{\infty}a_l\Psi^{(l)}_{+}
\end{split}
\end{equation}
and try to find the coefficients $a_l$ so that the combination $\Psi^{(0)}_{+}+\frac{1}{N}\Psi^{(1)}_{+}$ is a well defined CFT operator.\par
To make the bulk field well defined we need to cancel the pole at $Y=1$ in \eqref{zerothpsi+}. Also \eqref{hpsi+} has an ambigious $i\epsilon$ prescription from the branch cut for $Y\geq 1$. So we need to find the coeficients in \eqref{perturbativeeq} such that the branch cuts cancel after summing over $l$ while a surviving  pole at $Y=1$ cancel the problematic pole in \eqref{zerothpsi+}. Combining \eqref{hpsi+} and \eqref{perturbativeeq} we have
\begin{equation}\label{cpsi+}
\begin{split}
\left<\Psi^{(1)}_+\left(Z, x_1\right)\psi^{\dag}_+\left(x_2\right)j_{-}\left(x_3\right)\right>=\frac{Z^{\frac{1}{2}}{\chi_2\left(\chi_1-1\right)}}{\left(x_{23}^-\right)\left(x_{12}^+\right)\left(Z\right)^{2h}\left(\chi_1 -1\right)^{2h+2}}\left(\frac{\chi_1}{\chi_2}\right)^2\frac{1}{(Y-1)^2}f(Y)
\end{split}
\end{equation}
where,
\begin{equation}
\begin{split}
f(Y)=\sum_{l=0}^{\infty}a_l\hat{\alpha}^{(l)}(-Y)^{l+1}F\left(2h+l,l,2h+2l+2,Y\right)
\end{split}
\label{fy}
\end{equation}
Looking at what we need to cancel from (\ref{zerothpsi+bet}) it is convinient to parametrize $f(Y)$ as
\begin{equation}\label{104}
\begin{split}
f(Y)=\hat{\gamma}\left[2h-(2h-1)Y+(Y-1)^2g(Y)\right]
\end{split}
\end{equation}
So the corrected three point function is
\begin{equation}\label{correctedpsi+}
\begin{split}
\left<(\Psi^{(0)}_{+}+\frac{1}{N}\Psi^{(1)}_{+})\left(Z, x_1\right)\psi^{\dag}_+\left(x_2\right)j_{-}\left(x_3\right)\right>=\frac{\hat{\gamma}Z^{\frac{1}{2}}{\chi_2\left(\chi_1-1\right)}}{\left(x_{23}^-\right)\left(x_{12}^+\right)\left(Z\right)^{2h}\left(\chi_1 -1\right)^{2h+2}}\left(\frac{\chi_1}{\chi_2}\right)^2g(Y)
\end{split}
\end{equation}
Now just by examining this equation we can have some guesses. The $a_{l}$ should be such that  $f(Y)$  has no branch cuts at $Y>1$ and from equation (\ref{fy}) $f(Y) \rightarrow 0$  as $Y \rightarrow 0$. So the simplest possibility is that $f(Y)$ is a polynomial in $Y$, and thus so should $g(Y)$ be a polynomial. The simplest one that gives $f(Y) \rightarrow 0$  as $Y \rightarrow 0$ is $g(Y)=-2h$, which corresponds to $f(Y)=-2hY^2+(2h+1)Y$. This gives the same correlator we computed in (\ref{corrected_correlation_psipn}).

 This is similar to what happened in the scalar field case. Of course other options are fine as well for instance $g(Y)=-2h+\alpha Y$. However it is the simplest choice that actually corresponds to the solution of the bulk equation of motion we considered and the other conditions. The more general possibilities are related by some bulk field redefinition (as in the scalar case).
The correlator for the higher dimensional operator related to $\Psi_{-}$ from (\ref{psil_corr}) can be written as

\begin{equation}
\begin{split}
\left<\Psi^{(1)}_{-}\psi^{\dag}_{+}j_{-}\right>=\sum_{l} \frac{Z^{1/2}}{x_{23}^- Z^{2h+1}\left(\chi_1-1\right)^{2h+1}}&\bigg[(Y-1)b_{l}f^{(l)}_1(Y)+\chi_1b_{l}f^{(l)}_{2}(Y)+(1-Y)c_{l}f^{(l)}_3(Y)\\
&+\chi_1Yc_{l}f^{(l)}_3(Y)+(Y-1)c_{l}f^{(l)}_4(Y)\bigg]\left(\frac{\chi_1}{\chi_2}\right)\frac{1}{(Y-1)^2}
\end{split}
\end{equation}
where in the last line we have used $\frac{\chi_1}{\chi_2}=1+\chi_1Y-Y$.\\

The choice of $b_{l}$ and $c_{l}$ should be such as to cancel all the branch cuts in the hypergeometric functions and  to cancel the pole in 
\begin{equation}
\begin{split}
\langle\Psi^{(0)}_-\left(Z, x_1\right)\psi^{\dag}_+\left(x_2\right)j_{-}\left(x_3\right)\rangle=\frac{\tilde{\gamma_1}~Z^{\frac{1}{2}}\left(2h-(2h-1)\chi_2Y\right)}{(x_{23}^-)(Z)^{2h+1}(\chi_1 -1)^{2h+1}}\left(\frac{\chi_1}{\chi_2}\right)^2\frac{1}{(Y-1)^2}
\end{split}
\end{equation}
with $\tilde{\gamma_1}=\frac{iq}{4h\pi N}$.
This can be written as 
\begin{equation}
\begin{split}
\langle\Psi^{(0)}_-\left(Z, x_1\right)\psi^{\dag}_+\left(x_2\right)j_{-}\left(x_3\right)\rangle=\frac{\tilde{\gamma_1}~Z^{\frac{1}{2}}\left(2h(1-Y)+\chi_1Y\right)}{(x_{23}^-)(Z)^{2h+1}(\chi_1 -1)^{2h+1}}\left(\frac{\chi_1}{\chi_2}\right)\frac{1}{(Y-1)^2}.
\end{split}
\end{equation}

Now it is convinient to parametrize
\begin{equation}
\begin{split}
&\frac{1}{N}\sum_{l=0}^{\infty}\left[b_{l} f^{(l)}_1(Y)-c_{l} f^{(l)}_3(Y)+c_{l} f^{(l)}_4(Y)\right]=\tilde{\gamma_1}(2h+(Y-1)g(Y))\\
&\frac{1}{N}\sum_{l=0}^{\infty}\left[b_{l} f^{(l)}_2(Y)+Yc_{l} f^{(l)}_3(Y)\right]=\tilde{\gamma_1}(-Y+\tilde{g}(Y)(Y-1)^2)
\end{split}
\end{equation}
with this parametrization we have 
\begin{equation}
\begin{split}
\langle(\Psi^{(0)}_-\left(Z, x_1\right)+\frac{1}{N} \Psi^{(1)}_-\left(Z, x_1\right)) \psi^{\dag}_+\left(x_2\right)j_{-}\left(x_3\right)\rangle=\frac{\tilde{\gamma_1}~Z^{\frac{1}{2}}\left(g(Y)+\chi_{1}\tilde{g}(Y)\right)}{(x_{23}^-)(Z)^{2h+1}(1-\chi_1)^{2h+1}}\left(\frac{\chi_1}{\chi_2}\right)
\end{split}
\end{equation}

Now the absence of any branch cuts for $Y>1$ is simply satisfied by taking (guessing) $g(Y)$ and $\tilde{g}(Y)$ to be polynomials.
Since all the $f_{i}(Y) \rightarrow 0$ as $Y \rightarrow 0$ the simplest possibility is $g(Y)=2h$ and $\tilde{g}(Y)=0$. 
Then one finds that indeed we get the correlation function computed before in (\ref{corrected_correlation_psimn}).

\centerline{\bf Acknowledgements}
We would like to thank Sarthak Duary for collaboration in the early stages of this work.
G.L would like to thank D. Kabat for discussions. M.P would like to thank Sayantani Bhattacharyya for useful discussion and encouragement.
G.L is supported in part by the Israel Science Foundation under grant 447/17.
\appendix

\section{Computing 3-point function with $\Psi_{\pm}^{(0)}$}
In this section we derive the CFT correlators given in (\ref{oth:co_p}) and (\ref{oth:co_m}). 
We will assume that $2h$ is an integer.
We start with the three point correlator with a conserved current $j_{-}$ of dimension $\left(1,0\right)$, which is given by
\begin{equation}\label{0thorder3}
\begin{split}
\left<\psi_+\left(x_1\right)\psi^{\dag}_+\left(x_2\right)j_{-}\left(x_3\right)\right>=&\frac{\gamma_1}{\left(x_{12}^{-}\right)^{2h-1}\left(x_{12}^{+}\right)^{2h+1}\left(x_{13}^{-}\right)\left(x_{23}^{-}\right)}\\
=&\frac{\gamma_1}{\left(x_{23}^{-}\right)}\frac{\partial^{2h-2}_{x_2^-}\partial^{2h}_{x_2^+}}{\Gamma(2h-1)\Gamma(2h+1)}\frac{1}{x_{12}^{-}x_{12}^{+}x_{13}^{-}}
\end{split}
\end{equation}
with $\gamma_1=-i\frac{q}{2 \pi N}$. To compute $\langle\Psi^{(0)}_+\left(Z, x_1\right)\psi^{\dag}_+\left(x_2\right)j_{-}\left(x_3\right)\rangle$ we have to evaluate some derivatives acting on 
\begin{equation}
\begin{split}
\int_{{t^{\prime}}^2+{y^{\prime}}^2\leq Z^2}dt^{\prime}dy^{\prime}\left(\frac{Z^2-{y^{\prime}}^2-{t^{\prime}}^2}{Z}\right)^{\Delta-2-\frac{1}{2}}\frac{1}{\left(x_{12}^{-}-iy^{\prime}+t^{\prime}\right)\left(x_{12}^{+}+iy^{\prime}+t^{\prime}\right)\left(x_{13}^{-}-iy^{\prime}+t^{\prime}\right)}
\end{split}
\end{equation}
To evaluate the integral we change variable to $t^{\prime}=r\cos\theta,~y^{\prime}=r\sin\theta,~\alpha=e^{i\theta}$ to get
\begin{equation}
\begin{split}
\int_0^{Z}rdr\left(\frac{Z^2-r^2}{Z}\right)^{\Delta-2-\frac{1}{2}}\oint_{|\alpha|=1}\frac{\alpha d\alpha}{i\left(\alpha x_{12}^{-}+r\right)\left( x_{12}^{+}+r\alpha\right)\left(\alpha x_{13}^{-}+r\right)}
\end{split}
\end{equation}
The integration contour encloses the poles at $\alpha=-r/x_{12}^-$ and $\alpha=-r/x_{13}^-$, which gives
\begin{equation}
\begin{split}
\int_0^{Z}rdr\left(\frac{Z^2-r^2}{Z}\right)^{\Delta-2-\frac{1}{2}}\frac{2\pi x_{12}^+}{\left(r^2-x_{12}^+x_{12}^-\right)\left(r^2-x_{12}^+x_{13}^-\right)}
\end{split}
\end{equation}
Then the correlator for $\Psi^{(0)}_+$ can be evaluated by
\begin{equation}
\begin{split}
&\langle\Psi^{(0)}_+\left(Z, x_1\right)\psi^{\dag}_+\left(x_2\right)j_{-}\left(x_3\right)\rangle\\
&=\frac{2\gamma_1~(2h-1)Z^{\frac{1}{2}}}{\left(x_{23}^{-}\right)}\frac{\partial^{2h-2}_{x_2^-}\partial^{2h}_{x_2^+}}{\Gamma(2h-1)\Gamma(2h+1)}\int_0^{Z}rdr\left(\frac{Z^2-r^2}{Z}\right)^{2h-2}\frac{x_{12}^+}{\left(r^2-x_{12}^+x_{12}^-\right)\left(r^2-x_{12}^+x_{13}^-\right)}
\end{split}
\end{equation}
Taking derivative with respect to $x_{2}^-$ we have
\begin{equation}
\begin{split}
\frac{2(-1)^{2h}\gamma_1~(2h-1)Z^{\frac{1}{2}}}{\left(x_{23}^{-}\right)}\frac{\partial^{2h}_{x_2^+}}{\Gamma(2h+1)}\int_0^{Z}rdr\left(\frac{Z^2-r^2}{Z}\right)^{2h-2}\frac{\left(x_{12}^+\right)^{2h-1}}{\left(r^2-x_{12}^+x_{12}^-\right)^{2h-1}\left(r^2-x_{12}^+x_{13}^-\right)}
\end{split}
\end{equation}
Changing variable to $s=\frac{r2}{x_{12}^+}$ we get
\begin{equation}
\begin{split}
\frac{(-1)^{2h}\gamma_1~(2h-1)Z^{\frac{1}{2}}}{\left(x_{23}^{-}\right)}\frac{\partial^{2h}_{x_2^+}}{\Gamma(2h+1)}\int_0^{\frac{Z^2}{x_{12}^+}}ds\left(\frac{Z^2-x_{12}^+s}{Z}\right)^{2h-2}\frac{1}{\left(s-x_{12}^-\right)^{2h-1}\left(s-x_{13}^-\right)}
\end{split}
\end{equation}
which gives
\begin{equation}
\begin{split}
(-1)^{2h}\frac{\gamma_1~Z^{\frac{1}{2}}}{(2h)(x_{23}^-)}\frac{\partial}{\partial x_2^+}\left(\frac{(Z)^{2h}}{\left(Z^2-x_{12}^+x_{12}^-\right)^{2h-1}\left(Z^2-x_{12}^+x_{13}^-\right)}\right).
\end{split}
\end{equation}
The 3-point correlator can be rewritten as
\begin{equation}
\begin{split}
\left<\Psi^{(0)}_+\left(Z, x_1\right)\psi^{\dag}_+\left(x_2\right)j_{-}\left(x_3\right)\right>=\frac{iq}{4h\pi N}\frac{Z^{\frac{1}{2}}\left[(2h-1)\chi_1+\chi_2-2h\chi_1\chi_2\right]}{x_{23}^- x_{12}^+ Z^{2h}\left(\chi_1 -1\right)^{2h}\left(1-\chi_2\right)^2}
\end{split}
\end{equation}

where we have introduced the  combinations $\chi_1=\frac{x_{12}^+x_{12}^-}{Z^2}$ and $\chi_2=\frac{x_{12}^+x_{13}^-}{Z^2}$.\\
Finally the correlator can be written as
\begin{equation}
\begin{split}
\left<\Psi^{(0)}_+\left(Z, x_1\right)\psi^{\dag}_+\left(x_2\right)j_{-}\left(x_3\right)\right>=\frac{iq}{4h\pi N}\frac{Z^{\frac{1}{2}}\left[(2h-1)\chi_1+\chi_2-2h\chi_1\chi_2\right]}{x_{23}^- x_{12}^+ Z^{2h}\left(\chi_1 -1\right)^{2h+2}}\left(\frac{\chi_1}{\chi_2}\right)^2\frac{1}{(Y-1)^2}
\end{split}
\end{equation}
where we have introduced the combination
\begin{equation}
Y=\frac{\chi_2-\chi_1}{\chi_2\left(1-\chi_1\right)}
\end{equation}

Similarly, the correlator for $\Psi^{(0)}_-$ can be evaluated by
\begin{equation}
\begin{split}
&\langle\Psi^{(0)}_-\left(Z, x_1\right)\psi^{\dag}_+\left(x_2\right)j_{-}\left(x_3\right)\rangle\\
&=\frac{2\gamma_1~Z^{\frac{1}{2}}}{\left(x_{23}^{-}\right)}\frac{\partial_{x_1^-}\partial^{2h-2}_{x_2^-}\partial^{2h}_{x_2^+}}{\Gamma(2h-1)\Gamma(2h+1)}\int_0^{Z}rdr\left(\frac{Z^2-r^2}{Z}\right)^{2h-1}\frac{x_{12}^+}{\left(r^2-x_{12}^+x_{12}^-\right)\left(r^2-x_{12}^+x_{13}^-\right)}.
\end{split}
\end{equation} 
\unboldmath
Following similar steps as before we get 

\begin{equation}
\begin{split}
\left<\Psi^{(0)}_-\left(Z, x_1\right)\psi^{\dag}_+\left(x_2\right)j_{-}\left(x_3\right)\right>=\frac{iq}{4h\pi N}\frac{Z^{\frac{1}{2}}\left(2h-(2h-1)\chi_2Y\right)}{x_{23}^- Z^{2h+1}(\chi_1 -1)^{2h+1}}\left(\frac{\chi_1}{\chi_2}\right)^2\frac{1}{(Y-1)^2}.
\end{split}
\end{equation}

\section{Solving the equation of motion}

In this section we will obtain the bulk operator by solving bulk equation of motion, without assuming any knowledge (from the transformation under the modular Hamiltonian)  about the structure of the operators used to construct the bulk fermions.

Now, to cancel the ambiguous singularities in the correlator of \eqref{oth:co_p} and \eqref{oth:co_m}, we have to add a tower of higher-dimension double-trace operators to the bulk fields we have defined. The available ingredients are the operators of the form 
\begin{equation}
\begin{split}
\partial^{r}_{-}j_{-}\left(x^{-}\right)\partial^{s}_{-}\partial^{r+s}_{+}\psi_{+}\left(x^{+}, x^{-}\right),~~~r,s=0,1,2,\cdots
\end{split}
\end{equation}

The equation of motion for $\Psi_{\pm}^{(1)}$ are given by 

\begin{equation}\label{1storder_component}
\begin{split}
&\begin{pmatrix}{-Z\partial_Z\Psi^{(1)}_-+\left(1-m\right)\Psi^{(1)}_-+2Z\partial_-\Psi^{(1)}_+}\\\\ {-2Z\partial_+\Psi^{(1)}_-+Z\partial_Z\Psi^{(1)}_+-\left(1+m\right)\Psi^{(1)}_+}\end{pmatrix}=-2\frac{q}{k}Z\begin{pmatrix}ij_{-}\left(x^{-}\right)\Psi^{(0)}_{+} \\\\ 0 \end{pmatrix}
\end{split}
\end{equation}
Where  $\Psi^{(0)}_+$ can be given as 
\begin{equation}
\begin{split}
\Psi^{(0)}_+=\sum_{N=0}^{\infty}\frac{\Gamma\left(m+\frac{1}{2}\right)}{N!\Gamma\left(N+m+\frac{1}{2}\right)}Z^{2N+m+1}\partial_-^N\partial_+^N\psi_+\left(x^+,x^-\right).
\end{split}
\end{equation}
Now we will use the following notations for $\Psi^{(1)}_+$, $\Psi^{(1)}_-$ and $\Psi^{(0)}_+$
\begin{equation}
\begin{split}
&\Psi^{(1)}_+=\sum_{M,N=0}^{\infty}C^{(+)}_{MN}Z^{2M+2N+m+3}\partial_-^{M}j_-(x^-)\partial_-^{N}\partial_+^{M+N+1}\psi_+(x^+,x^-)\\
&\Psi^{(1)}_-=\sum_{M,N=0}^{\infty}C^{(-)}_{MN}Z^{2M+2N+m+2}\partial_-^{M}j_-(x^-)\partial_-^{N}\partial_+^{M+N}\psi_+(x^+,x^-)\\
&\Psi^{(0)}_+=\sum_{N=0}^{\infty}C^{(0)}_NZ^{2N+m+1}\partial_-^N\partial_+^N\psi_+\left(x^+,x^-\right)
\end{split}
\end{equation}
Now we will start solving \eqref{1storder_component} step by step.\\
The bottom part of the equation gives
\begin{equation}\label{befinal}
\begin{split}
\sum_{M,N=0}^{\infty}\left[\left(M+N+1\right)C^{(+)}_{MN}-C^{(-)}_{MN}\right]Z^{2M+2N+m+3}\partial_-^Mj_-(x^-)\partial_-^N\partial_+^{M+N+1}\psi_+\left(x^+,x^-\right)=0
\end{split}
\end{equation}
We will get the following equation form \eqref{befinal}
\begin{equation}
\begin{split}
\boxed{\left(M+N+1\right)C^{(+)}_{MN}-C^{(-)}_{MN}=0}
\end{split}
\end{equation}
The top part of the equation will give the following equation
\begin{equation}\label{tefinal}
\begin{split}
&-\sum_{M,N=0}^{\infty}C^{(-)}_{MN}\left(2M+2N+2m+1\right)Z^{2M+2N+m+2}\partial_-^Mj_-(x^-)\partial_-^N\partial_+^{M+N}\psi_+(x^+,x^-)\\
&+2\sum_{M,N=0}^{\infty}C^{(+)}_{MN}Z^{2M+2N+m+4}\bigg[\partial_-^{M+1}j_-(x^-)\partial_-^N\partial_+^{M+N+1}\psi_+(x^+,x^-)+\partial_-^{M}j_-(x^-)\partial_-^{N+1}\partial_+^{M+N+1}\psi_+(x^+,x^-)\bigg]\\
=&-2i\frac{q}{k}\sum_{N=0}^{\infty}C^{(0)}_NZ^{2N+m+2}j_-(x^-)\partial_-^N\partial_+^N\psi_+\left(x^+,x^-\right)
\end{split}
\end{equation}
\normalsize
From that one of the equation we will get as
\begin{equation}
\begin{split}
\boxed{C^{(-)}_{00}=\frac{2iq}{k}\frac{1}{2m+1}C^{(0)}_0}
\end{split}
\end{equation}
Another equation we will get as
\begin{equation}
\begin{split}
&-\sum_{R=0}^{\infty}C^{(-)}_{0,R+1}\left(2R+2m+3\right)Z^{2R+m+4}j_-(x^-)\partial_-^{R+1}\partial_+^{R+1}\psi_+(x^+,x^-)\\
&+2\sum_{R=0}^{\infty}C^{(+)}_{0,R}Z^{2R+m+4}j_-(x^-)\partial_-^{R+1}\partial_+^{R+1}\psi_+(x^+,x^-)\\
=&-2i\frac{q}{k}\sum_{R=0}^{\infty}C^{(0)}_{R+1}Z^{2R+m+4}j_-(x^-)\partial_-^{R+1}\partial_+^{R+1}\psi_+(x^+,x^-)
\end{split}
\end{equation}
Hence, from this we will get the following equation
\begin{equation}
\begin{split}
\boxed{-\left(2R+2m+3\right)C^{(-)}_{0,R+1}+2C^{(+)}_{0,R}=-2i\frac{q}{k}C^{(0)}_{R+1}}
\end{split}
\end{equation}
Another equation we will get
\begin{equation}
\begin{split}
&-\sum_{R=0}^{\infty}C^{(-)}_{R+1,0}\left(2R+2m+3\right)Z^{2R+m+4}\partial_-^{R+1}j_-(x^-)\partial_+^{R+1}\psi_+(x^+,x^-)\\
&+2\sum_{R=0}^{\infty}C^{(+)}_{R,0}Z^{2R+m+4}\partial_-^{R+1}j_-(x^-)\partial_+^{R+1}\psi_+(x^+,x^-)
=0
\end{split}
\end{equation}
which implies the following equation
\begin{equation}
\begin{split}
\boxed{-\left(2R+2m+3\right)C^{(-)}_{R+1,0}+2C^{(+)}_{R,0}=0}
\end{split}
\end{equation}
Another equation we will get as
\begin{equation}
\begin{split}
&-\sum_{R,T=0}^{\infty}C^{(-)}_{R+1,T+1}\left(2R+2T+2m+5\right)Z^{2R+2T+m+6}\partial_-^{R+1}j_-(x^-)\partial_-^{T+1}\partial_+^{R+T+2}\psi_+(x^+,x^-)\\
&+2\sum_{R=0}^{\infty}C^{(+)}_{R,T+1}Z^{2R+2T+m+6}\partial_-^{R+1}j_-(x^-)\partial_-^{T+1}\partial_+^{R+T+2}\psi_+(x^+,x^-)\\
&+2\sum_{R,T=0}^{\infty}C^{(+)}_{R+1,T}Z^{2R+2T+m+6}\partial_-^{R+1}j_-(x^-)\partial_-^{T+1}\partial_+^{R+T+2}\psi_+(x^+,x^-)=0\\
\end{split}
\end{equation}
From which we will get the following equation
\begin{equation}
\begin{split}
\boxed{-\left(2R+2T+2m+5\right)C^{(-)}_{R+1,T+1}+2C^{(+)}_{R,T+1}+2C^{(+)}_{R+1,T}=0}
\end{split}
\end{equation}
Now if we solve the system of equations written in boxes using Mathematica we will get The $C_{MN}^{\pm}$ of equations (\ref{cnm+}) and (\ref{cnm-}).

\section{Computing the building blocks}

We have seen that the bulk operator is built from certain smearings of $J_{\psi}^{(l)}$ which is a primary with conformal dimensions $\left(h+l+1,h+\frac{1}{2}\right)$ where 
\begin{equation}
J_{\psi}^{(l)}=\sum_{s+r=l} d_{r,s}\partial^{r}_{-}j_{-}\partial^{s}_{-}\psi_{+}, \  \ \ d_{r,s}=\frac{(-1)^{r}}{\Gamma(r+1) \Gamma(s+1)\Gamma (r+2)\Gamma (s+2h)}.
\end{equation}

The particular smearings are given by
\begin{equation}
\Psi^{(1)}_{+}=\frac{Z^{1/2}}{\pi} \sum_{l=0}^{\infty} a_{l} \int_{t'^2+y^2\leq Z^2} \left( \frac{Z^2-t'^2-y^2}{Z}\right)^{2h+2l}\partial_{+}^{l+1}J_{\psi}^{(l)}(t+t',x+iy)
\end{equation}
\begin{eqnarray}
\Psi^{(1)}_{-} &=& \frac{Z^{1/2}}{\pi} \sum_{l=0}^{\infty} b_{l} \int_{t'^2+y^2\leq Z^2} \left( \frac{Z^2-t'^2-y^2}{Z}\right)^{2h+2l-1}\partial_{+}^{l}J_{\psi}^{(l)}(t+t',x+iy)\nonumber \\
&+& \frac{Z^{1/2}}{\pi} \sum_{l=0}^{\infty} c_{l} \int_{t'^2+y^2\leq Z^2} \left( \frac{Z^2-t'^2-y^2}{Z}\right)^{2h+2l+1}\partial_{-}\partial_{+}^{l+1}J_{\psi}^{(l)}(t+t',x+iy)\nonumber\\
\end{eqnarray}

Let us define the building blocks
\begin{equation}
\Psi^{(l)}_{+}=\frac{Z^{1/2}}{\pi}  \int_{t'^2+y^2\leq Z^2} \left( \frac{Z^2-t'^2-y^2}{Z}\right)^{2h+2l}\partial_{+}^{l+1}J_{\psi}^{(l)}(t+t',x+iy),
\end{equation}
\begin{equation}
\Psi^{(l,1)}_{-} = \frac{Z^{1/2}}{\pi} \int_{t'^2+y^2\leq Z^2} \left( \frac{Z^2-t'^2-y^2}{Z}\right)^{2h+2l-1}\partial_{+}^{l}J_{\psi}^{(l)}(t+t',x+iy),
\end{equation}
\begin{equation}
\Psi^{(l,2)}_{-}= \frac{Z^{1/2}}{\pi} \int_{t'^2+y^2\leq Z^2} \left( \frac{Z^2-t'^2-y^2}{Z}\right)^{2h+2l+1}\partial_{-}\partial_{+}^{l+1}J_{\psi}^{(l)}(t+t',x+iy).
 \end{equation}

\subsection{Computing $\langle\Psi^{(l,i)}_-\left(Z, x_1\right)\psi^{\dag}_+\left(x_2\right)j_{-}\left(x_3\right)\rangle$}

We start with the three point function
\begin{equation}
\begin{split}
\left<J^{(l)}_{\psi}\left(x_1\right)\psi^{\dag}_{+}\left(x_2\right)j_{-}\left(x_3\right)\right>=&\frac{\alpha^{(l)}\left(x_{23}^-\right)^l}{\left(x_{12}^{-}\right)^{2h+l}\left(x_{12}^{+}\right)^{2h+1}\left(x_{13}^-\right)^{l+2}}\\
=&\alpha^{(l)}\left(x_{23}^-\right)^l\frac{\partial_{x_2^-}^{2h+l-1}\partial_{x_2^+}^{2h}\partial_{x_3^-}^{l+1}}{\Gamma(2h+l)\Gamma(2h+1)\Gamma(l+2)}\frac{1}{x_{12}^-x_{12}^+x_{13}^-}
\end{split}
\end{equation}
Using this the bulk correlation of $\Psi^{(l,1)}_-$  is given by
\begin{equation}
\begin{split}
&\langle\Psi^{(l,1)}_{-}\left(x_1,Z\right)\psi^{\dag}_{+}\left(x_2\right)j_{-}\left(x_3\right)\rangle=\frac{\alpha^{(l)}\left(x_{23}^-\right)^lZ^{\frac{1}{2}}}{\pi\Gamma(l+2)\Gamma(2h+l)\Gamma(2h+1)}\partial_{x_3^-}^{l+1}\partial_{x_2^-}^{2h+l-1}\partial_{x_2^+}^{2h}\partial_{x_1^+}^{l}\\
&\int_{{t^{\prime}}^2+{y^{\prime}}^2\leq Z^2}dt^{\prime}dy^{\prime}\left(\frac{Z^2-{y^{\prime}}^2-{t^{\prime}}^2}{Z}\right)^{2h+2l-1}\frac{1}{\left(x_{12}^{-}+\bar{w}\right)\left(x_{12}^{+}+w\right)\left(x_{13}^{-}+\bar{w}\right)}
\end{split}
\end{equation}
where, $w=t+t^{\prime}+iy^{\prime}$, $\bar{w}=t^{\prime}-iy^{\prime}$. We set $t^{\prime}=r~\cos\theta$, $y^{\prime}=r~\sin\theta$ and let $\alpha=e^{i\theta}$ so that $w=r~\alpha$, $\bar{w}=r/\alpha$ and then the integration measure becomes
\begin{equation}
\begin{split}
\int_{{t^{\prime}}^2+{y^{\prime}}^2\leq Z^2}dt^{\prime}dy^{\prime}=\int_0^Zr~dr\oint_{|\alpha|=1}\frac{d\alpha}{i\alpha}
\end{split}
\end{equation}
Now the contour integral over $\alpha$ leads to
\begin{equation}
\begin{split}
&\langle\Psi^{(l,1)}_{-}\left(x_1,Z\right)\psi^{\dag}_{+}\left(x_2\right)j_{-}\left(x_3\right)\rangle=\frac{\alpha^{(l)}\left(x_{23}^-\right)^lZ^{\frac{1}{2}}}{\pi\Gamma(l+2)\Gamma(2h+l)\Gamma(2h+1)}\partial_{x_3^-}^{l+1}\partial_{x_2^-}^{2h+l-1}\partial_{x_2^+}^{2h}\partial_{x_1^+}^{l}\\
&\int_0^Zrdr\left(\frac{Z^2-r^2}{Z}\right)^{2h+2l-1}\frac{2\pi~x_{12}^+}{\left(r^2-x_{12}^+x_{12}^-\right)\left(r^2-x_{12}^+x_{13}^-\right)}
\end{split}
\end{equation}
We first act the $x_2^{-}$ and $x_3^{-}$ derivatives keeping one $x_2^{-}$ derivative outside
\begin{equation}
\begin{split}
&\left<\Psi^{(l,1)}_{-}\psi^{\dag}_{+}j_{-}\right>\\
&=\frac{\alpha^{(l)}(-1)^{2h+1}\left(x_{23}^-\right)^lZ^{\frac{1}{2}}}{(2h)!(2h+l-1)}\partial_{x_2^-}~\partial_{x_2^+}^{2h}\partial_{x_1^+}^{l}\int_0^Zrdr\left(\frac{Z^2-r^2}{Z}\right)^{2h+2l-1}\frac{2~{\left(x_{12}^+\right)}^{2h+2l}}{\left(r^2-x_{12}^+x_{12}^-\right)^{2h+l-1}\left(r^2-x_{12}^+x_{13}^-\right)^{l+2}}
\end{split}
\end{equation}
Now changing the integration variable from $r$ to $\widetilde{s}=\frac{r^2}{x_{12}^+}$ and then we obtain
\begin{equation}
\begin{split}
&\left<\Psi^{(l,1)}_{-}\psi^{\dag}_{+}j_{-}\right>\\
&=\frac{\alpha^{(l)}(-1)^{2h+1}\left(x_{23}^-\right)^lZ^{\frac{1}{2}}}{(2h)!(2h+l-1)}\partial_{x_2^-}\partial_{x_2^+}^{2h}\partial_{x_1^+}^{l}\int_0^{Z^2/x_{12}^+}d\widetilde{s}\left(\frac{Z^2-x_{12}^+~\widetilde{s}}{Z}\right)^{2h+2l-1}\frac{1}{\left(\widetilde{s}-x_{12}^-\right)^{2h+l-1}\left(\widetilde{s}-x_{13}^-\right)^{l+2}}
\end{split}
\end{equation}
Then acting on the $x_2^{+}$ and $x_1^{+}$ derivatives we have\footnote{This is only true for $l>0$ and the $l=0$ case has to be dealt separately. However the final answer is actually correct for the case $l=0$ as well.} 
\begin{equation}
\begin{split}
\left<\Psi^{(l,1)}_{-}\psi^{\dag}_{+}j_{-}\right>=&\frac{(-1)^{2h+l+1}\alpha^{(l)}\left(x_{23}^-\right)^lZ^{\frac{1}{2}}(2h+2l-1)!}{(2h)!(2h+l-1)(l-1)!}\\
&\partial_{x_2^-}\int_0^{Z^2/x_{12}^+}d\widetilde{s}\left(\frac{Z^2-x_{12}^+~\widetilde{s}}{Z}\right)^{l-1}\left(\frac{\widetilde{s}}{Z}\right)^{2h+l}\frac{1}{\left(\widetilde{s}-x_{12}^-\right)^{2h+l-1}\left(\widetilde{s}-x_{13}^-\right)^{l+2}}
\end{split}
\end{equation}
Now we change the integration variable to $t=\frac{x_{12}^{+}\widetilde{s}}{Z^2}$, which gives
\begin{equation}
\begin{split}
&\left<\Psi^{(l,1)}_{-}\psi^{\dag}_{+}j_{-}\right>\\
&=\frac{(-1)^l\alpha^{(l)}(2h+2l-1)!\left(x_{12}^+\right)^{l}\left(x_{23}^-\right)^lZ^{\frac{1}{2}}}{(2h)!(2h+l-1)(l-1)!\left(Z\right)^{2h+2l+1}}\partial_{x_2^-}\left[\frac{1}{(\chi_1)^{2h+l-1}\left(\chi_2\right)^{l+2}}\int_0^1dt\frac{(t)^{2h+l}(1-t)^{l-1}}{\left(1-\frac{t}{\chi_1}\right)^{2h+l-1}\left(1-\frac{t}{\chi_2}\right)^{l+2}}\right]
\end{split}
\end{equation}
where $\chi_1=\frac{x_{12}^+x_{12}^-}{Z^2}$ and $\chi_2=\frac{x_{12}^+x_{13}^-}{Z^2}$.

Then we will use the following hypergeometric identity
\begin{equation}
\begin{split}
\int_0^1du\frac{u^{\alpha-1}(1-u)^{\gamma-\alpha-1}}{(1-u~x)^{\beta}(1-u~y)^{\beta^{\prime}}}&=\frac{\Gamma(\alpha)\Gamma(\gamma-\alpha)}{\Gamma(\gamma)}F_1\left(\alpha,\beta,\beta^{\prime},\gamma,x,y\right)\\
&=\frac{\Gamma(\alpha)\Gamma(\gamma-\alpha)}{\Gamma(\gamma)}\left(1-y\right)^{-\alpha}F\left(\alpha,\beta,\beta+\beta^{\prime},\frac{x-y}{1-y}\right)
\end{split}
\end{equation}
where, $F_1$ is Appell's $F_1$ function and the identity is true when $\gamma=\beta+\beta^{\prime}$.

Then the correlator can be written as
\begin{equation}
\begin{split}
\left<\Psi^{(l,1)}_{-}\psi^{\dag}_{+}j_{-}\right>=&\frac{{\alpha}^{(l)}(\chi_1-1)^l(Y)^lZ^{\frac{1}{2}}}{(Z)^{2h+1}}\frac{(2h+l)!}{(2h)!(2h+2l)(2h+l-1)}\\
&\partial_{x_2^-}\left[\frac{1}{(\chi_1-1)^{2h+l-1}(\chi_2-1)^2}F(2h+l-1,l,2h+2l+1,Y)\right]
\end{split}
\end{equation}
where in the last line we have also used the following identity
\begin{equation}
F\left(a,b,c,z\right)=(1-z)^{-b}F\left(b,c-a,c,\frac{z}{z-1}\right)
\end{equation}
After taking the $x_2^-$ derivative we can write the correlator as
\begin{equation}
\begin{split}
\left<\Psi^{(l,1)}_{-}\psi^{\dag}_{+}j_{-}\right>=&\frac{\alpha^{(l)}(2h+l)!Z^{\frac{1}{2}}}{(2h)!(2h+2l)(2h+l-1)\left(Z\right)^{2h+1}\left(x_{23}^-\right)}\\
&\bigg[\frac{1}{(\chi_1-1)^{2h}(\chi_2-1)}(Y)^{l+1}\frac{d}{dY}F(2h+l-1,l,2h+2l+1,Y)\\
&-\frac{(2h+l-1)\chi_2}{(\chi_1-1)^{2h-1}(\chi_2-1)^2}(Y)^{l+1}F(2h+l-1,l,2h+2l+1,Y)\bigg]
\end{split}
\end{equation}
which again can be written as
\begin{equation}
\begin{split}
\left<\Psi^{(l,1)}_{-}\psi^{\dag}_{+}j_{-}\right>&=\frac{\left(-\alpha^{(l)}_{b}\right)Z^{\frac{1}{2}}}{(x_{23}^-)(Z)^{2h+1}(\chi_1-1)^{2h+1}}\left(\frac{\chi_1}{\chi_2}\right)\frac{1}{(Y-1)}\left(Y\right)^{l+1}\frac{d}{dY}F(2h+l-1,l,2h+2l+1,Y)\\
&+\frac{\left(-\alpha^{(l)}_{b}\right)(2h+l-1)Z^{\frac{1}{2}}(\chi_2)}{(x_{23}^-)(Z)^{2h+1}(\chi_1-1)^{2h+1}}\left(\frac{\chi_1}{\chi_2}\right)^2\frac{1}{(Y-1)^2}(Y)^{l+1}F(2h+l-1,l,2h+2l+1,Y)
\end{split}
\end{equation}
where,
\begin{equation}
\begin{split}
\alpha^{(l)}_{b}=\frac{\alpha^{(l)}(2h+l)!}{(2h)!(2h+2l)(2h+l-1)}
\end{split}
\end{equation}

Similarly one can compute the 3-point function with $\Psi^{(l,2)}$. Following similar steps one arrives at 
\begin{equation}
\begin{split}
\left<\Psi^{(l,2)}_{-}\psi^{\dag}_{+}j_{-}\right>&=-\frac{\Tilde\alpha^{(l)}_c~(l+2)Z^{\frac{1}{2}}}{(x_{23}^-)(Z)^{2h+1}(\chi_1-1)^{2h+1}}\left(\frac{\chi_1}{\chi_2}\right)^2\frac{1}{(Y-1)^2}(Y)^{l+1}F\left(2h+l,l+1,2h+2l+3,Y\right)\\
&+\frac{\Tilde\alpha^{(l)}_c~(2h+l)Z^{\frac{1}{2}}}{(x_{23}^-)(Z)^{2h+1}(\chi_1-1)^{2h+1}}\left(\frac{\chi_1}{\chi_2}\right)\frac{1}{(Y-1)}(Y)^{l+1}F\left(2h+l+1,l+1,2h+2l+3,Y\right)
\end{split}
\end{equation}
where
\begin{equation}
\begin{split}
\Tilde\alpha^{(l)}_c=\frac{\alpha^{(l)}(2h+l+1)!}{(2h)!(2h+2l+2)}
\end{split}
\end{equation}
Adding all contributions we get 
\begin{equation}\label{psil_corr}
\begin{split}
\left<\Psi^{(1)}_{-}\psi^{\dag}_{+}j_{-}\right>=& \sum_{l} \frac{b_{l} Z^{1/2}}{(x_{23}^-)(Z)^{2h+1}\left(\chi_1-1\right)^{2h+1}}\left(\frac{\chi_1}{\chi_2}\right)\frac{1}{(Y-1)}f_1^{(l)}(Y)\\
&+ \sum_{l}\frac{b_{l} Z^{1/2}\left(\chi_2\right)}{(x_{23}^-)(Z)^{2h+1}\left(\chi_1-1\right)^{2h+1}}\left(\frac{\chi_1}{\chi_2}\right)^2\frac{1}{(Y-1)^2}f_2^{(l)}(Y)\\
&+ \sum_{l}\frac{c_{l}Z^{1/2}}{(x_{23}^-)(Z)^{2h+1}\left(\chi_1-1\right)^{2h+1}}\left(\frac{\chi_1}{\chi_2}\right)^2\frac{1}{(Y-1)^2}f_3^{(l)}(Y)\\
&+ \sum_{l}\frac{c_{l}Z^{1/2}}{(x_{23}^-)(Z)^{2h+1}\left(\chi_1-1\right)^{2h+1}}\left(\frac{\chi_1}{\chi_2}\right)\frac{1}{(Y-1)}f_4^{(l)}(Y)\\
\end{split}
\end{equation}
where,
\begin{equation}
\begin{split}
&f^{(l)}_1(Y)=\left(-\alpha^{(l)}_b\right)(Y)^{l+1}\frac{d}{dY}F\left(2h+l-1,l,2h+2l+1,Y\right)\\
&f^{(l)}_2(Y)=\left(-\alpha^{(l)}_b\right)(2h+l-1)(Y)^{l+1}F\left(2h+l-1,l,2h+2l+1,Y\right)\\
&f^{(l)}_3(Y)=\Tilde\alpha^{(l)}_c(l+2)(Y)^{l+1}F\left(2h+l,l+1,2h+2l+3,Y\right)\\
&f^{(l)}_4(Y)=\Tilde\alpha^{(l)}_c(2h+l)(Y)^{l+1}F\left(2h+l+1,l+1,2h+2l+3,Y\right)
\end{split}
\end{equation}

 \subsection{Computing $\langle\Psi^{(l)}_+\left(Z, x_1\right)\psi^{\dag}_+\left(x_2\right)j_{-}\left(x_3\right)\rangle$}

Like in the previous subsection we will start from the 3-point correlator
\begin{equation}
\begin{split}
&\left<J^{(l)}_{\psi}\left(x_1\right)\psi^{\dag}_{+}\left(x_2\right)j_{-}\left(x_3\right)\right>=\frac{\alpha^{(l)}\left(x_{23}^-\right)^l}{\left(x_{12}^{-}\right)^{2h+l}\left(x_{12}^{+}\right)^{2h+1}\left(x_{13}^-\right)^{l+2}}
\end{split}
\end{equation}
Using this the correlation of $\Psi^{(l)}_+$ with $\psi^{\dag}_+$ and $j_{-}$ is given by
\begin{equation}
\begin{split}
&\left<\Psi^{(l)}_{+}\left(x_1,Z\right)\psi^{\dag}_{+}\left(x_2\right)j_{-}\left(x_3\right)\right>=\frac{\alpha^{(l)}Z^{1/2}\left(x_{23}^-\right)^l}{\pi\Gamma(l+2)\Gamma(2h+l)\Gamma(2h+1)}\partial_{x_3^-}^{l+1}\partial_{x_2^-}^{2h+l-1}\partial_{x_2^+}^{2h}\partial_{x_1^+}^{l+1}\\
&\int_{{t^{\prime}}^2+{y^{\prime}}^2\leq Z^2}dt^{\prime}dy^{\prime}\left(\frac{Z^2-{y^{\prime}}^2-{t^{\prime}}^2}{Z}\right)^{2h+2l}\frac{1}{\left(x_{12}^{-}+\bar{w}\right)\left(x_{12}^{+}+w\right)\left(x_{13}^{-}+\bar{w}\right)}
\end{split}
\end{equation}
where, $w=t+t^{\prime}+iy^{\prime}$, $\bar{w}=t^{\prime}-iy^{\prime}$. 
Following similar steps as before we get 
\begin{equation}
\begin{split}
\left<\Psi^{(l)}_{+}\psi^{\dag}_{+}j_{-}\right>=\frac{\hat{\alpha}^{(l)}~Z^{1/2}~\left[\chi_2\chi_1-\chi_2\right]}{\left(x_{23}^-\right)\left(x_{12}^+\right)\left(Z\right)^{2h}\left(\chi_1 -1\right)^{2h+2}}\left(\frac{\chi_1}{\chi_2}\right)^2\frac{1}{(Y-1)^2}(-Y)^{l+1}F\left(2h+l,l,2h+2l+2,Y\right)
\end{split}
\label{psil+}
\end{equation}
where,
\begin{equation}
\hat{\alpha}^{(l)}=\frac{(-1)^{l+1}\alpha^{(l)}\Gamma(2h+l+2)}{(2h+2l+1)\Gamma(2h+1)}.
\end{equation}


\end{document}